\documentclass[aps,showpacs,superscriptaddress,nofootinbib]{revtex4}
\usepackage{epsfig,rotating}
\usepackage{amsmath,amssymb}
\usepackage{dsfont}

\numberwithin{equation}{section}
\newcommand{\dbarp}{\frac{d^3p}{(2\pi)^3}}
\newcommand{\dbarq}{\frac{d^3q}{(2\pi)^3}}
\newcommand{\vx}{\vec{x}}
\newcommand{\vp}{\vec{p}}
\newcommand{\vq}{\vec{q}}
\newcommand{\vk}{\vec{k}}
\newcommand{\la}{\lambda}
\newcommand{\om}{\omega}
\newcommand{\p}{\omega_p}
\newcommand{\ga}{\gamma}

\newcommand{\uvk}{\widehat{\bf{k}}}

\newcommand{\be}{\begin{equation}}
\newcommand{\ee}{\end{equation}}
\newcommand{\bea}{\begin{eqnarray}}
\newcommand{\eea}{\end{eqnarray}}
\newcommand{\no}{\noindent}

\begin{document}

\title{Neutrino Oscillations in the Early Universe: \\ A Real-Time
Formulation}
\author{\bf C. M. Ho}
\email{cmho@phyast.pitt.edu}\affiliation{  Department of Physics and
Astronomy, University of Pittsburgh, Pittsburgh, Pennsylvania 15260,
USA }
\author{\bf D. Boyanovsky }
\email{boyan@pitt.edu}\affiliation{  Department of Physics and
Astronomy, University of Pittsburgh, Pittsburgh, Pennsylvania 15260,
USA }\affiliation{Observatoire de Paris, LERMA, Laboratoire
Associ\'e au CNRS UMR 8112,
 \\61, Avenue de l'Observatoire, 75014 Paris, France. }
\author{\bf H. J. de Vega }\email{devega@lpthe.jussieu.fr}
\affiliation{LPTHE, Universit\'e Pierre et Marie Curie (Paris VI) et
Denis Diderot (Paris VII), Laboratoire Associ\'e au CNRS UMR 7589,
Tour 24, 5\`eme. \'etage, 4, Place Jussieu, 75252 Paris, Cedex 05,
France} \affiliation{Observatoire de Paris, LERMA, Laboratoire
Associ\'e au CNRS UMR 8112,
 \\61, Avenue de l'Observatoire, 75014 Paris, France. }

\date{\today}

\begin{abstract}
Neutrino oscillations in the early Universe prior to the epoch of
primordial nucleosynthesis is studied by implementing real time
non-equilibrium field theory methods. We focus on two flavors of
Dirac neutrinos, however, the formulation is general. We obtain the
equations of motion for neutrino wavepackets of either chirality and
helicity in the plasma  allowing for CP asymmetry. Contributions
non-local in space-time to the self-energy dominate over the
asymmetry for $T \gtrsim 3-5\,\mathrm{MeV}$ if the lepton and
neutrino asymmetries are of the same order as the baryon asymmetry.
We find a new contribution which cannot be interpreted as the usual
effective potential. The mixing angles and dispersion relations in
the medium depend on \emph{helicity}. We find that resonant
transitions are possible in the temperature range $ 10 \lesssim T
\ll 100\,\mathrm{MeV} $. Near a resonance in the mixing angle, the
oscillation time scale in the medium as compared to the vacuum is
\emph{slowed-down} substantially for small vacuum mixing angle. The
time scale of oscillations \emph{speeds-up} for off resonance high
energy neutrinos for which the mixing angle becomes vanishingly
small. The equations of motion reduce to the familiar oscillation
formulae for negative helicity ultrarelativistic neutrinos, but
include consistently both the \emph{mixing angle and the oscillation
frequencies in the medium}. These equations of motion also allow to
study the dynamics of right handed as well as positive helicity
neutrinos.
\end{abstract}

 \maketitle

\section{Introduction}\label{sec:intro}

Neutrinos have taken center stage in particle physics and have
become a bridge between astrophysics, cosmology, particle physics
and nuclear physics \cite{book1,book2,book3,raffelt,bah}.

A wealth of experimental data have confirmed that neutrinos are
massive and that different flavors of neutrinos can mix and
oscillate \cite{panta,giunti,smirnov, bilenky, haxton,
grimus,gouvea}  thus  providing indisputable evidence for
\emph{new physics} beyond the Standard Model.

Neutrinos  play a fundamental role in cosmology and astrophysics,
and it is now widely accepted that resonant flavor oscillations
due to the  MSW effect in the sun  provide a concrete explanation
to the solar neutrino problem \cite{MSWI, MSWII,bah,bethe}.

Neutrino oscillations in extreme conditions of temperature and
density are  an important aspect of Big Bang Nucleosynthesis (BBN)
and in the generation of the lepton asymmetry in the early
Universe\cite{fuller,dolgov,haxton,raffelt,kirilova}, as well as
in the physics of core collapse supernovae\cite{SN,panta,bethe},
and the formation, evolution and cooling of neutron
stars\cite{prakash,reddy,yakovlev}.

While accelerator and reactor experiments  measure the
\emph{difference} of the
 neutrino masses, high precision cosmological observations of the
 cosmic microwave background by WMAP combined with large scale
 structure  suggest that the \emph{sum} of the masses of all
neutrino species is bound to be smaller than   $1 eV$ \cite{WMAP}.

An important aspect of neutrino oscillations is lepton number
violation, leading to the  suggestion that the baryon asymmetry may
actually originate in  the lepton sector and the  proposal
 that leptogenesis can be the main mechanism that explains the
cosmological baryon asymmetry \cite{fukugita,yanagida,buch}.

Neutrino propagation in a cold medium has been first studied in ref.
\cite{MSWI} wherein the refractive index of electron neutrinos was
computed. The early studies of neutrino propagation focused on the
neutrino dispersion relations and damping rates in the temperature
regime relevant for stellar evolution or big bang nucleosynthesis
\cite{notzold,dolgov}. Since then, the work has been extended to
include leptons, neutrinos and nucleons in the medium
\cite{dolivoDR}. The matter effects of neutrino oscillations in the
early universe has been investigated in \cite{barbieri,enqvist}.

Since the original study of neutrino oscillations in the
sun\cite{MSWI,MSWII}, neutrino oscillations are typically studied
within the single particle quantum mechanical formulation. For two
flavors the evolution Hamiltonian is simply that for a two-state
system, where the off-diagonal terms lead to the mixing and
oscillation phenomena. The medium properties are input in this
formulation after computing the contributions from charged and
neutral currents in a medium with leptons, neutrinos and hadrons or
quarks. A conceptually similar approach underlies the kinetic
treatment of oscillations in the early Universe wherein the dynamics
is studied from the time evolution of a density matrix that
generalizes the \emph{single particle} description but that does
\emph{not} generally account for the subtle aspects of flavor Fock
states addressed in ref\cite{DanMan}, which introduce a hierarchy of
time scales. Recently\cite{kine} a set of generalized
\emph{semiclassical} Boltzmann equations for the single particle
distribution functions, supposedly applicable to neutrino transport
in core-collapse supernovae have been proposed, but where the mixing
term must be obtained \emph{separately} from the underlying field
theory. For a thorough discussion of the kinetic approach to mixing
and relaxation and the approximations involved, the reader is
referred to\cite{dolgov,sigl,kine}.

The dynamics of neutrino mixing in the presence of a background of
neutrinos requires in general a full non-linear treatment, which has
so far been studied within a self-consistent single particle
framework\cite{self} in the form of approximate kinetic equations
for the reduced density matrix\cite{dolgov,sigl}. Such study
revealed a wealth of novel non-linear phenomena such as
self-synchronization\cite{self}. An approximate treatment of
background neutrinos within the framework of \emph{equilibrium}
finite temperature field theory has also been proposed in
ref.\cite{JCDOlivo}. However, a consistent treatment must
necessarily rely on a non-linear kinetic description, which has not
yet been developed in the full field theory framework.

 A  calculation of the neutrino dispersion relations in a hot and dense medium implementing
the techniques of quantum field theory at finite temperature was
provided in ref.\cite{notzold}. This treatment was extended in
ref.\cite{JCDOlivo} to study the propagation of mixed neutrinos of
negative helicity in  a neutrino background, up to lowest order in
$g^2/M^2_W$.  In ref.\cite{panta2} the quantum fields for Dirac
neutrinos propagating in a cold but dense medium and the dispersion
relation for both chirality components  were obtained, again
considering the medium corrections to the dispersion relations up to
order $g^2/M^2_W$.

The study of dynamics of neutrino mixing in the literature is
mostly carried out within the framework of a single particle
description, wherein the  dynamical evolution is described in
terms of an effective Hamiltonian for either a two or three level
system (depending the number of flavors). However, a single
particle formulation is inadequate in a hot and or dense medium
where collective many body effects may be predominant. The main
point in the above discussion is to highlight that there is a leap
in the current approach to study neutrino oscillations in a
medium: the result of a quantum field theory calculation of the
index of refraction or effective potentials in the medium is input
into a single particle quantum mechanical description of
oscillations and mixing based on Bloch-type equations.

We have previously reported on a study of  oscillations in  a hot
and/or  dense neutrino gas directly from the underlying quantum
field theory\cite{DanMan} in free field theory. Such study, even
at the free field level revealed  quantum interference phenomena
and subtle many body aspects responsible for a hierarchy of time
scales that cannot be captured within the single particle
description. In particular, the subtle aspects of the Fock
representation of the distribution function as well as
interference and coherence phenomena, leading to widely different
time scales   has not been fully included in the treatments in
ref.\cite{dolgov,sigl,kine}.

More recently, novel \emph{collective} neutrino excitations in the
\emph{standard model} (namely without masses and mixing) near the
critical temperature were studied\cite{boya}. Collective phenomena
requires a systematic and consistent treatment implementing the
methods of quantum field theory at finite temperature and density.
Furthermore, the \emph{real time} dynamics of mixing and
oscillations requires a non-equilibrium formulation of quantum field
theory specially suited to study the real time evolution as an
\emph{initial value problem}\cite{ctp}-\cite{nosfermions}.

There are at least four fundamental reasons to study neutrinos in
the early Universe at a deeper level: i) neutrino mixing may be
\emph{the} mechanism by which baryogenesis is a result of
leptogenesis\cite{fukugita,yanagida,buch}, ii) big bang
nucleosynthesis is particularly sensitive to the spectrum and
 oscillations of neutrinos\cite{dolgov2,kirilova,barbieri}, iii)
just like the cosmic microwave background, there is
 a cosmic neutrino background left over from the big bang, and iv) neutrinos
masses and mixing are the most
 clear experimental  confirmation  of physics beyond the standard
 model. All of these reasons warrant a complete quantum field
 theory study of neutrinos in the early Universe.

\vspace{2mm}

{\bf The goals of this article:} \\

The full quantum field theory treatment of neutrino mixing in hot
and or dense media has not yet received the same level of attention
as the more familiar single particle treatment, which however, is
not suited when  collective many body phenomena become relevant as
is typically the case in extreme environments.

Previous quantum field theory
studies\cite{notzold,JCDOlivo,panta2,dolgov} address either the
dispersion relations \emph{or} mixing phenomena under restrictive
approximations  to lowest order in $g^2/M^2_W$.

 In this article we provide a systematic quantum field theory study   of neutrino propagation
and oscillations in the   early Universe \emph{directly in real
time}. Because in the early Universe the lepton asymmetries are
expected to be typically of the same order of the baryon asymmetry
$\eta_B/\eta_{\gamma} \sim 10^{-9}$ a consistent description of
neutrino propagation and oscillations requires to include
corrections non-local in space-time of order $g^2/M^4_W$ in the
dispersion relations\cite{notzold} \emph{and mixing angles}. We
focus our study on the case of two flavors of Dirac neutrinos, taken
to be the electron and muon neutrinos, this study can be generalized
to more flavors or to Majorana-Dirac mass matrices without any
conceptual difficulty. Our main goals are: {\bf i )} To provide a
systematic and consistent study of the real time dynamics of
neutrino oscillation and mixing directly in quantum field theory in
conditions of temperature and lepton/neutrino asymmetries applicable
to the early Universe prior to the nucleosynthesis era.  This is
achieved  by formulating an initial value problem via linear
response and implementing real time field theory methods at finite
temperature and density. {\bf ii)} To obtain the dispersion
relations and in-medium mixing angles including the \emph{non local}
contributions from the neutrino self-energies up to order
$g^2/M^4_W$. The one-loop self-energy in expanded  to lowest order
in $(\omega,k)/M_W$. We find a new contribution which cannot be
interpreted as the usual effective potential.  These contributions
are necessary since the typical asymmetries in the early Universe
are very small and these non-local (in space-time) contributions can
be of the same order or larger than the local contributions.{\bf
iii)} To obtain the in-medium Dirac spinors for both helicities  and
study the evolution of oscillations and mixing for \emph{both}
helicity components directly in real time. {\bf iv)} Two different
temperature regimes are studied in detail: i) $m_e \ll T \ll m_\mu$,
ii) $m_e,m_\mu \ll T \ll M_W$. The first regime is just prior to big
bang nucleosynthesis. Lepton and hadron (proton and neutrons in
nuclear statistical equilibrium) or quark asymmetries are included
in the one-loop self-energy. We assess in detail the temperature and
energy regime for which a resonance in the   mixing angle is
available in the medium. The second temperature regime is above the
QCD phase transition and we include two flavor of (light quarks)
with their respective asymmetries. In this regime the mixing angle
becomes small. In both cases we also study the mixing and
oscillations of positive helicity as well as right handed neutrinos,
which are typically neglected in the literature. We also obtain the
loop corrections to the \emph{oscillation frequencies} thereby
providing  a complete description of oscillation and mixing that
includes corrections to both the mixing angle and the oscillation
frequencies. {\bf v)}  We obtain general oscillation formulae
derived directly from the real time evolution in quantum field
theory. These formulae reveal the limit in which the usual quantum
mechanical single particle description is reliable  as well as the
corrections to them.

\textbf{Main approximations:}  Since our study relies on a one-loop
self-energy computation including leptons and neutrinos, the
inclusion of a neutrino background must necessarily imply some
approximations for consistency.

We do not \emph{yet} consider absorptive contributions, which in the
temperature regime studied here are of two-loop order, postponing
the study of the interplay between oscillations and relaxation to a
forthcoming article.

Since we obtain the non-local (in space-time) contributions from the
one-loop self energy we must address the issue of the neutrino
propagators in the neutral current contributions. Because of mixing,
the neutrino propagator in the flavor basis, in which the weak
interactions are diagonal,  does not correspond to the propagation
of mass eigenstates and in principle the non-equilibrium propagators
obtained in ref.\cite{DanMan} must be used. The question of
equilibration of a neutrino gas with mixing is one of time scales:
the weak interactions are diagonal in the flavor basis, therefore
weak processes tend to equilibrate \emph{flavor} neutrinos with  a
typical weak interaction relaxation rate at high temperature
\cite{notzold} $ \Gamma \sim G^2_F \; T^5$. Oscillations, on the
other hand mix flavors and tend to redistribute flavor neutrinos
into
 mass eigenstates of energy $ E $ on a time scale
$ \tau_{osc} \sim E/\delta M^2 $. Combined fitting of the solar
and KamLAND data yield \cite{kamland} $ |\delta M^2 | \approx
7.9\times 10^{-5}\,(eV)^2$, therefore considering $ E \sim T $, we
find $ \Gamma \; \tau_{osc} \sim 10 \; (0.1\,T/\textrm{MeV})^6 $.
This comparison of time scales suggests that for $ T \gtrsim
10\,\textrm{MeV} $ neutrinos are equilibrated \emph{as flavor
eigenstates}. Flavor eigenstates created at local weak interaction
vertices will reach thermal equilibrium on time scales far shorter
than those required for oscillations into mass eigenstates for
temperatures larger than $ \sim 10\,\textrm{MeV} $. Since in a
loop integral the typical momenta are of order $ T $, and
\emph{assuming} the validity of this estimate, we   consider the
neutrino propagators in the neutral current self-energy loop to be
diagonal in the flavor basis, massless  and in thermal
equilibrium.

For temperatures $T \lesssim 10 \,\mathrm{MeV}$ and certainly
below freeze out $T < 1\,\mathrm{MeV}$ a full kinetic description
that includes oscillations and expansion\cite{barbieri,dolgov} is
required. The study of the kinetic equations will be the subject
of forthcoming work. In this article we restrict our study to the
temperature regime $T \gtrsim 10 \, \mathrm{MeV}$.

We also \emph{assume} that the lepton and neutrino asymmetries are
of the same order as the baryon asymmetry, namely  $ L_i =
(n_i-\bar{n}_i)/n_\gamma \sim 10^{-9} $. For a relativistic species
the asymmetry is proportional to $ \xi_i \; (1+\xi^2_i/\pi^2) $ with
$ \xi_i \equiv \mu_i/T $, therefore under this assumption $ \xi_i
\sim 10^{-9} $ and we can safely neglect the contribution to the
chemical potential in the non-local (in space-time) terms of order $
g^2/M^4_W $.

{\bf Brief summary of main results:} The main results obtained in
this article are: {\bf i)} We obtain the equation of motion in real
time for initially prepared wavepackets of neutrinos. Both
chiralities and helicities are treated on equal footing. The
self-energy is expanded up to lowest order in the frequency and
momentum yielding \emph{non-local} contributions to the equations of
motion. These happen to be \emph{larger} than those from the lepton
and neutrino asymmetries for $T \gtrsim 5-10 \,\mathrm{MeV}$. {\bf
ii)} We studied two different temperature regimes: $m_e \ll T \ll
m_\mu$ within which we show  that there is the possibility of
resonant oscillations of test neutrinos, and $m_e, m_\mu \ll T \ll
M_W$ within which the mixing angle for active neutrinos effectively
vanishes.  For $T \sim 10\,\mathrm{MeV}$ resonant flavor
oscillations occur for neutrino energies in the few $ \mathrm{MeV}$
range. {\bf iii)} Mixing angles    in the medium not only depend on
energy but also on \emph{helicity}. The dispersion relations of
propagating neutrinos  in the medium also depend  on helicity.
Assuming that the lepton and quark asymmetries are of the same order
as the baryon asymmetry in the early Universe, the non-local  (in
space-time) terms in the self-energies dominate over the asymmetry
for typical energies of neutrinos in the plasma for $T \gtrsim 3-5\,
\mathrm{MeV}$.  {\bf iv)}The oscillation time scale in the medium is
\emph{slowed-down } near the resonance, becoming substantially
\emph{longer} than in the vacuum for small vacuum mixing angle. For
high energy neutrinos off-resonance the mixing angle becomes
vanishingly small and the oscillation time scale \emph{speeds-up} as
compared to the vacuum. {\bf v)} The equations of motion reduce to
the familiar oscillation formulae for ultrarelativistic negative
helicity neutrinos, but they consistently include the mixing angles
and the \emph{oscillation frequencies} in the medium. We obtain
general oscillation formula for either chirality and helicity. These
equations also describe the  oscillation  dynamics of \emph{right
handed} neutrinos, which, while suppressed as consequence of the
small masses,  are not sterile.

The article is organized as follows: in section \ref{sec:effective}
we obtain the equations of motion for initially prepared neutrino
wavepackets by implementing the methods of non-equilibrium field
theory and linear response. In section \ref{sec:self-energy} we
obtain the one-loop self-energy contributions from charged and
neutral currents. Section \ref{sec:diagonal} is devoted to obtaining
the dispersion relations, mixing angles and oscillation time scales
in the medium and a study of the possibility of resonances. In
section \ref{sec:laplace} we study the real time evolution of
neutrino wavepackets as an initial value problem. Section
\ref{conclu} presents our conclusions, summarizes our results and
presents some conjectures and further questions. The detailed
calculation of the self-energy is presented in an appendix.

\section{Effective Dirac equation for neutrino propagation in a
medium}\label{sec:effective}

The propagation of a neutrino in a medium is determined by the
effective Dirac equation which includes the self-energy corrections.
Its solution yields the real time evolution as \emph{an initial
value problem}. The correct framework to study the dynamics is the
real time formulation of field theory in terms of the
closed-time-path integral\cite{ctp}-\cite{nosfermions}. In this
section we implement this method combined with linear response to
obtain the effective equation of motion for an \emph{expectation
value} of the neutrino field. The main concept in this approach is
the following, consider coupling an external c-number Grassman
source to the neutrino field and switching this source adiabatically
up to time $t=0$. This source induces an expectation value of the
neutrino field, after switching-off the external source at $t=0$,
the expectation value evolves in time as a solution of the effective
Dirac equation in the medium with the initial condition determined
by the source term.

The main ingredient in this program is the retarded self-energy
which enters in the effective Dirac equation. The real-time
formulation of field theory directly leads to causal and retarded
equations of motion. It is important to highlight the difference
with the S-matrix approach which describes transition amplitudes
from in to out states, the real time formulation yields the
equations of motion for an \emph{expectation} value and these are
fully causal\cite{ctp}-\cite{nosfermions}.

The self-energy is obtained in the unitary gauge in which only the
correct physical degrees of freedom contribute and is manifestly
unitary\cite{peskin}. Previous calculations of the neutrino
self-energy in covariant gauges (one of which is the unitary gauge)
have proven that although the self-energy does depend on the gauge
parameter, the dispersion relations are gauge-invariant
\cite{dolivoDR}.

As mentioned above we restrict our discussion to the case of two
 flavors of Dirac neutrinos, namely the electron  and  muon neutrinos. The
subtle $CP$ violating phases associated with the case of three
active neutrinos will not be considered here. However, the method
can be  generalized to three active neutrinos, sterile neutrinos
or even Majorana neutrinos without any conceptual difficulty and
will be postponed for further discussion elsewhere.

For Dirac neutrinos, mixing and oscillations can be implemented by
a minimal extension of the standard model  adding a Dirac mass
matrix to the standard model Lagrangian which is off-diagonal in
the flavor basis. The relevant part of the Lagrangian density is
given by

\be\label{LSM} \mathcal{L} = \mathcal{L}^0_{\nu} +
\mathcal{L}^0_W+\mathcal{L}^0_Z +
\mathcal{L}_{CC}+\mathcal{L}_{NC},\ee

\noindent where $\mathcal{L}^0_{\nu}$ is the free field  neutrino
Lagrangian minimally modified to include a Dirac mass matrix

 \be\label{FFnuL} \mathcal{L}^0_{\nu} = \overline{\nu}_a
\left(i {\not\!{\partial}}\,\delta_{ab}-M_{ab}\right) \nu_b \ee

\noindent with $a, b$ being the flavor indexes. For two flavors of
Dirac neutrinos the mass matrix $M_{ab}$ is given by
\be \label{massmatrix} \mathds{M}=\left(%
\begin{array}{cc}
  m_{ee} & m_{e\mu} \\
  m_{e\mu} & m_{\mu\mu} \\
\end{array}%
\right) \; , \ee $\mathcal{L}^0_{W,Z}$ are the free field lagrangian
densities for the vector bosons in the unitary gauge, namely \bea
\mathcal{L}^0_W &=& -\frac12
\left(\partial_{\mu}W^+_{\nu}-\partial_{\nu}W^+_{\mu}
\right)\left(\partial^{\mu}W^{-\,\nu}-\partial^{\nu}W^{-\,\mu}
\right)+ M^2_W \; W^+_{\mu} \; W^{-\,\mu} \; , \label{LW}\\
\mathcal{L}^0_Z & = & -\frac{1}{4}
\left(\partial_{\mu}Z_{\nu}-\partial_{\nu}Z_{\mu}
\right)\left(\partial^{\mu}Z^{\nu}-\partial^{\nu}Z^{\mu} \right)+
\frac12M^2_Z \; Z_{\mu} \; Z^{\mu} \; , \label{LZ} \eea \noindent
and the charged and neutral current interaction lagrangian densities
are given by \be \mathcal{L}_{CC} = \frac{g}{\sqrt2} \left[
\overline{\nu}_a \; \gamma^\mu \;  L  \;  l_a \;  W^+_{\mu} +
\overline{l}_a \gamma^\mu \;  L \;  \nu_a \;  W^-_{\mu} \right] \;
,\label{LCC} \ee \be \mathcal{L}_{NC} = \frac{g}{2 \cos \theta_w}
\left[ \overline{\nu}_a \; \gamma^\mu \;  L \;  \nu_a \;  Z_{\mu} +
\overline{f}_a \; \gamma^\mu \;  (g^V_a-g^A_a\,\gamma^5) \;  f_a \;
Z_{\mu} \right]\label{LNC} \; . \ee \noindent where
$L=(1-\gamma^5)/2$ is the left-handed chiral projection operator,
$g^{V,A}$ are the vector and axial vector couplings for quarks and
leptons, $l$ stands for leptons and $f$ generically for the fermion
species with neutral current interactions.

For two flavors, the diagonalization of the free field Dirac
Lagrangian for neutrinos, (\ref{FFnuL}) is achieved by a unitary
transformation to mass eigenstates. Considering, flavor and mass
 doublets  respectively
$$    \left(
         \begin{array}{c}
           \nu_{e} \\
           \nu_{\mu  } \\
         \end{array}
       \right) \quad ,  \quad \left(
         \begin{array}{c}
           \nu_{1} \\
           \nu_{2  } \\
         \end{array}
       \right) \; , $$
        related by    unitary transformation
       \be  \left(
         \begin{array}{c}
           \nu_{e} \\
           \nu_{\mu  } \\
         \end{array}
       \right) = U  \left(
         \begin{array}{c}
           \nu_{1} \\
           \nu_{2  } \\
         \end{array}
       \right)  \; , \label{unitrafo}
\ee
 with the unitary transformation given by the $2\times 2$ matrix
\be \label{vacrot} U = \left(
          \begin{array}{cc}
            \cos\theta  & \sin \theta  \\
            -\sin \theta  & \cos \theta  \\
          \end{array}
        \right)  \; ,
\ee \noindent where $\theta$ is the \emph{vacuum} mixing angle.

In the basis of mass eigenstates $ (\nu_1,\nu_2) $ the mass matrix
$M_{ab}$ becomes diagonal
$$ \left(
          \begin{array}{cc}
            M_1  & 0  \\
            0  & M_2 \\
          \end{array}
        \right)  \; .
$$
The elements $ m_{ee}, m_{\mu\mu} $ and $ m_{e\mu} $ in the mass
matrix (\ref{massmatrix}) are related to the vacuum mixing angle $
\theta $ and  masses of the propagating mass eigenstates $ M_1 $ and
$ M_2$ as follows \bea m_{ee}=C^2 \;  M_1+S^2 \;
M_2~;~~m_{\mu\mu}=S^2 \;  M_1+C^2 \; M_2~;~~m_{e\mu}=-(M_1-M_2) \; C
\; S \; , \eea \no where $C=\cos\theta$ and $S=\sin\theta$.

For later convenience, we  introduce \be \overline{M}=
\frac{M_1+M_2}{2} ~~;~~ \delta M^2 = M^2_1-M^2_2 \label{mbardm} \ee
The current value for the average of the vacuum  masses obtained by
WMAP\cite{WMAP} and oscillation parameters from the combined fitting
of the solar and KamLAND data are \cite{kamland}: \be \overline{M}
\approx 0.25 \,(eV)~~;~~|\delta M^2| \approx 7.9\times
10^{-5}\,(eV)^2~;~~\tan^2\theta \approx 0.40 \; \,. \ee For these
values of the masses and more generally if there is an almost
degeneracy in the hierarchy of neutrino masses the ratio \be
\label{small}\frac{|\delta M^2|}{\overline{M}^{\,2}} \ll 1 \; . \ee
The smallness of this ratio in the nearly degenerate case will lead
to important simplifications.

Our goal is to obtain the effective Dirac equation for neutrinos
propagating in the medium and extract the in-medium mixing angles,
propagation frequencies and the wave functions of the propagating
modes in the medium. The real-time effective Dirac equation in the
medium is derived from linear response by  implementing the methods
of non-equilibrium quantum field theory described in
\cite{nosfermions}.

 Following this approach, we introduce an external
Grassmann-valued source that couples linearly to the neutrino field
via the lagrangian density \be \mathcal{L}_S = \overline{\nu}_a  \;
\eta_a + \overline{\eta}_a \; \nu_a \; . \label{Lsource} \ee
\noindent whence the total lagrangian density is given by
$\mathcal{L}+\mathcal{L}_S$. The external source induces an
expectation value for the neutrino field which, in turn, obeys the
effective equation of motion with self-energy modifications from the
thermal medium \cite{nosfermions}.

To study the dynamics of the system, it is the \emph{expectation
values} rather than the in-out S-matrix elements that are
necessary\cite{Jordan}. This requires a generating function for the
real-time correlation functions. Denoting generically by $\Phi$ the
fields (fermions, or gauge bosons), a path integral representation
of this generating functional is given by \be\label{Z}
\mathcal{Z}[j^+,j^-] = \int D\Phi^+ \, D\Phi^- e^{i \int
(\mathcal{L}[\Phi^+,j^+]-\mathcal{L}[\Phi^-,j^-])} \; , \ee \no and
the path integrations over the fields $\Phi^{\pm}$ will be taken
along the forward $(+)$ and backward $(-)$ time branches, in the
presence of the sources $j^{\pm}$ \cite{ctp}-\cite{nosfermions}.
Here, the sources $j^{\pm}$ are coupled linearly to the fields
$\Phi^{\pm}$ and thus the real-time correlation functions can be
obtained from the functional derivatives of this generating
functional with respect to these sources. Functional derivatives
with respect to $j^{+}$ and $j^{-}$ give the time-ordered and
anti-time-ordered correlation functions respectively.

The sources $j^{\pm}$ are introduced to compute the real-time
correlation functions and will be set to zero after the
calculations. However, the external Grassman source $\eta$ and hence
the expectation value induced to the neutrino field will remain the
same along both time branches $(\pm)$. For further discussions on
the general method, see references
\cite{ctp,disip,tadpole,nosfermions}.

The equation of motion for the expectation value of the neutrino
field induced by the external Grassman source is derived by shifting
the field \be \nu^{\pm}_a = \psi_a + \Psi^{\pm}_a \quad , \quad
\psi_a = \langle \nu^{\pm}_a \rangle  \quad , \quad \langle
\Psi^{\pm}_a \rangle =0 \; , \label{shift} \ee \noindent and
imposing $\langle \Psi^{\pm}_a \rangle =0$ order by order in the
perturbation theory \cite{disip,tadpole,nosfermions}. Carrying out
this implementation up to   one-loop order, we find the following
equation of motion \be
\left(i\not\!{\partial}\,\delta_{ab}-M_{ab}+\Sigma^{tad}_{ab} \;
L\right)\,\psi_b (\vx,t) + \int d^3 x' \int dt' \;
\Sigma^{ret}_{ab}(\vx-\vx',t-t') \; \psi_b(\vx',t') = -
\eta_a(\vx,t), \label{eqnofmot} \ee
\begin{figure}
\begin{center}
\includegraphics[height=3in,width=4in,keepaspectratio=true]{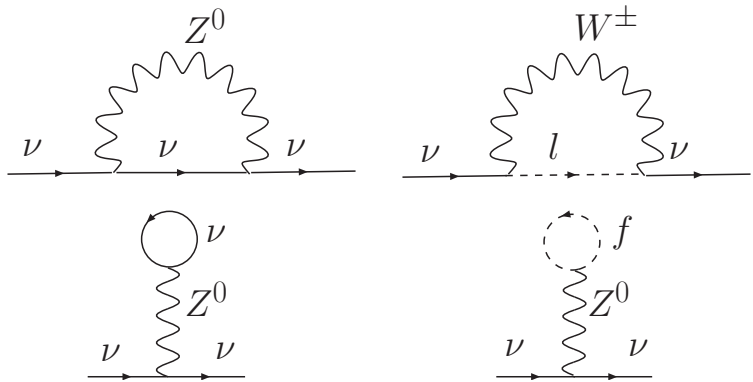}
\caption{One-loop diagrams contributing to the neutrino
self-energy.} \label{neusigma}
\end{center}
\end{figure}
\noindent where $ \Sigma^{ret}_{ab}(\vx-\vx',t-t') $ is the
real-time retarded self-energy given by the exchange one-loop
contributions, the first two diagrams  displayed in Fig.
\ref{neusigma} and $ \Sigma^{tad}_{ab} \; L $ is the tadpole
contribution from the last two diagrams in Fig. \ref{neusigma}. The
expectation value of the neutrino field in the medium describes a
beam or wave-packet of test neutrinos, namely these are neutrinos
that are injected in the medium, for example by the decay of a
neutron or any other heavy particle, but have not (yet) thermalized
with the plasma. This is precisely the manner in which linear
response leads to a study of real-time phenomena.

Due to the translational invariance of the medium in thermal
equilibrium, the retarded self-energy is simply a function of
$\vx-\vx'$ and $t-t'$. Hence,  the equation can be written in
frequency and momentum space by introducing the space-time Fourier
transform of the expectation value and the source \be \psi_a (\vx,t)
= \frac{1}{\sqrt{V}}\sum_{\vk}\int d\omega \; \psi_a (\vk,\omega) \;
e^{i\vk\cdot\vx} \; e^{-i\omega\,t} \quad , \quad  \eta_a(\vx,t) =
\frac{1}{\sqrt{V}}\sum_{\vk}\int d\omega  \; \eta_a (\vk,\omega) \;
e^{i\vk\cdot\vx} \; e^{-i\omega\,t}.\label{FTs} \ee \no Furthermore,
due to the rotational invariance of the thermal medium, implies that
all   physical quantities  depend  on $|\vec{k}|\equiv k$.

We have argued in the introduction that for temperatures $T
\gtrsim 10 \, \mathrm{MeV}$ the relaxation via the weak
interaction is faster than the time scales of (vacuum)
oscillations.  The validation of this assumption requires a deeper
study of the interplay between neutrino oscillations and damping
processes in the medium including the two-loop self-energy, which
will be the subject of a forthcoming article. \emph{Assuming} the
validity of this estimate, the neutrinos in the loop are in
thermal equilibrium as flavor eigenstates and at the temperatures
of interest we consider them massless. Under this assumption, the
self-energy is diagonal in the flavor basis

As a result, the effective Dirac equation for neutrino oscillations
in the medium is obtained as \be \left[\left(\gamma^0\,\omega-
\vec{\gamma}\cdot \vk\right)\delta_{ab}
-M_{ab}+\Sigma^{tad}_{ab}\,L+ \Sigma_{ab }(\om,k)\,L\right]\,\psi_b
(\omega,k)=-\eta_a(\omega,k)\, , \label{EqMotFT} \ee \no with \be
\Sigma(\om,k)=\Sigma_W(\omega,k)+ \Sigma_Z(\omega,k)~~;~~
\Sigma^{tad} = \Sigma^{\nu_e\,tad}+\Sigma^{\nu_\mu\,tad}+
\Sigma^{f\,tad} \; , \ee \no where $\Sigma^{f\,tad}$ is the total
tadpole contribution from all fermions other than neutrinos in the
loop.

The  details of the calculation of the self-energy  are presented in
the Appendix.

\section{One-Loop Self-Energies}\label{sec:self-energy}

\subsection{Charged and Neutral Currents}

From Eqs.(\ref{retSWdisp})-(\ref{imparts}) in the Appendix, we found
that the self-energies $\Sigma_{W,Z}(\omega,k)$ can be written in
the dispersive form \be\Sigma_{W,Z}(\omega,k) = \int
\frac{dk_0}{\pi} \;
\frac{\textit{Im}\,{\Sigma}_W(k_0,k)}{k_0-\omega-i\epsilon}.
\label{retSEfin2} \ee \no with \bea
&&\textit{Im}\,{\Sigma}_W(k_0,k)=\frac{g^2\pi}{2} \int \dbarp
\frac{1}{4 \; W_q \; \p} \Bigg\{ \left[
1-N_f(\p)+N_b(W_q)\right]\not\!{Q}(\vp,\vq)\,
\delta(k_0-\p-W_q) +\nonumber \\
&&+\left[ 1-\bar{N}_f(\p)+N_b(W_q)\right]\not\!{Q}(-\vp,-\vq)\,
\delta(k_0+\p+W_q) + \left[N_f(\p)+N_b(W_q)
\right]\not\!{Q}(\vp,-\vq)\,
\delta(k_0-\p+W_q)+\nonumber \\
&&+\left[\bar{N}_f(\p)+N_b(W_q) \right]\not\!{Q}(-\vp,\vq)\,
\delta(k_0+\p-W_q) \Bigg\} \; ,\label{ImsigW} \eea \noindent where
we have defined \bea && Q^{\mu}(\vp,\vq) =  p^\mu+ 2 \; q^\mu  \;
\frac{W_q\,\p-
\vq\cdot\vp}{M^2_W} \; , \label{bigQ} \\
&& q^\mu  =  (\,W_q\,,\vk-\vp\,) \quad , \quad \p=
\sqrt{|\vec{p}\,|^2+m_f^2}\quad , \quad
 W_q  =  \sqrt{|\,\vk-\vp\,\,|^2+M^2_W} \; .
\nonumber \eea The corresponding contribution from neutral currents
can be obtained from the above expression by setting \be
\frac{g}{\sqrt2} \rightarrow \frac{g}{2\cos\theta_w} \quad , \quad
M_W\rightarrow M_Z= \frac{M_W}{\cos\theta_w} \; ,\label{repla} \ee
\no with $\theta_w$ being the Weinberg angle.

In the limit $T\ll M_{W,Z}$, the abundance of vector bosons is
exponentially suppressed, hence we neglect the terms that feature
  $ N_{b}(W_q) $.  The imaginary part of the one-loop self-energy
  vanishes on the neutrino mass shell at one-loop level. A non-vanishing
    damping rate (non-vanishing imaginary part of the self-energy at the
neutrino mass shell) at temperatures $ T\ll   M_W $ arises at
two-loop level. Thus we focus solely on $
\textit{Re}\,{\Sigma}_W(k_0,k) $ when studying the dispersion
relation and propagation of neutrinos in the medium.

The form of $\textit{Im}\,{\Sigma}_W(k_0,k)$ suggests that
$\textit{Re}\,{\Sigma}_W(\om,k)$ can be written as \be
\textit{Re}\,{\Sigma}_W(\omega,k)=\gamma^0 \; \sigma^0_W(\omega,
k)-\vec\gamma\cdot \widehat{\bf{k}} \;
\sigma^1_W(\omega,k),\label{ReSelf} \ee \noindent where
$\sigma^0_W(\omega,k)$ and $\sigma^1_W(\omega,k)$ can be obtained by
taking traces on both sides.

Dropping the $T=0$ part and using the dispersive representation
(\ref{retSEfin2}), we find that for any fermion $f$ in the loop,
\bea \label{sig0W} {\sigma}^0_W(\omega,k) = -\frac{g^2}{2} \int
\dbarp \frac{1}{4 \; W_q \; \p}&\Bigg\{&
N_f(\p)\left[\frac{{Q}_{0}(\vp,\vq)}{W_q+\p-\omega}
+\frac{{Q}_{0}(\vp,-\vq)}{W_q-\p+\omega)} \right] \nonumber \\&&-
\bar{N}_f(\p)\left[\frac{{Q}_{0}(\vp,\vq)}{W_q+\p+\omega}
+\frac{{Q}_{0}(\vp,-\vq)}{W_q-\p-\omega}\right]\,\,\,\Bigg\} \; ,
\eea \bea \label{sig1W} {\sigma}^1_W(\omega,k) = -\frac{g^2}{2} \int
\dbarp \frac{1}{4 \; W_q \; \p}&\Bigg\{&
N_f(\p)\left[\frac{\uvk\cdot\vec{Q}(\vp,\vq)}{W_q+\p-\omega}
+\frac{\uvk\cdot\vec{Q}(\vp,-\vq)}{W_q-\p+\omega} \right] \nonumber
\\&&+\bar{N}_f(\p)\left[\frac{\uvk\cdot\vec{Q}(\vp,\vq)}{W_q+\p+\omega}
+\frac{\uvk\cdot\vec{Q}(\vp,-\vq)}{W_q-\p-\omega}\right]\,\,\,\Bigg\}
\; . \eea In the thermalized medium with temperature  $T $, the
dominant loop momenta is of order $p\sim T$, therefore we neglect
the neutrino masses since $T \gtrsim 3-5\, \mathrm{MeV}$. The
self-energy is expanded in a power series in the external frequency
and momentum $\omega,k$, we refer to terms of the form
$\omega/M_W~;~k/M_W$ as non-local  terms (in space-time) since they
represent gradient expansions in configuration space. Furthermore,
we  neglect the contribution from leptons with masses $m_f \gg T$
since these will be exponentially suppressed, but we calculate the
self-energies up the order $(g^2/M_{W,Z}^4)(m_f/T)^2$ for leptons
with masses $m_f \ll T$ and all higher order terms will be dropped.
A straightforward but lengthy calculation gives \bea
{\sigma}^0_W(\omega,k)&=& -\frac{3 \; G_F}{\sqrt2}(n_f-n_{\bar{f}})+
\frac{7 \; \pi^2}{15 \; \sqrt2}\frac{G_F \;
\om \; T^4}{M_W^2} \; , \nonumber \\
{\sigma}^1_W(\omega,k)&=& -\frac{7 \; \pi^2}{45 \; \sqrt2} \;
\frac{G_F \;  k \; T^4}{M_W^2}\left[1-\frac{30}{7 \;
\pi^2}\left(\frac{m_f}{T}\right)^2\right] \; , \label{sigmaW} \eea
\no where $G_F= \sqrt2  \; \frac{g^2}{8  \; M_W^2} $ is the Fermi
constant, and $ n_f-n_{\bar{f}} $ is the particle-antiparticle
number density difference
 for any fermion $f$  defined as
\be n_f-n_{\bar{f}}= 2\int\,\frac{d^{3}p}{(2\pi)^3}
\left[N_f(\p)-\bar{N}_f(\p)\right] \; . \ee The contribution to $
\sigma^1_W(\om,k) $ of order $ g^2/M_W^2 $ vanishes as a consequence
of the isotropy of the equilibrium distribution functions. In
calculating the non-local (in space-time) terms proportional to
$\omega/M_W~;~k/M_W$ we have neglected the chemical potentials under
the assumption that all asymmetries (leptons and neutrinos) are of
the same order as the baryon asymmetry, in which case $\mu/T \sim
10^{-9}$ where $\mu$ is the chemical potential for the corresponding
species.

In ref.\cite{notzold} the equivalent to $\sigma^0$ is quoted  as the
coefficient $b_L$, but no equivalent to $\sigma^1$ was provided
there, this is a difference between our results and those of
ref.\cite{notzold}. The contribution $\sigma^1(\omega,k)$ is
\emph{new} and it \emph{cannot} be identified with an ``effective
potential'' (which is proportional to $\gamma^0$) and it will be a
source of helicity dependence on frequencies and mixing angles which
has not been appreciated in the literature.

To obtain the corresponding expressions for the neutral current
interactions, we can simply apply the replacement eq.(\ref{repla}).
The tadpole contributions have been previously obtained in
ref.\cite{notzold} where we refer the reader for more details. The
result of the tadpole diagrams is summarized by the expression
 \be
\Sigma^{f\,tad}=\frac{2 \; G_F \; g^V}{\sqrt2} \; (n_f-n_{\bar{f}})
\; \gamma^0 \; , \ee

\no where the coefficient $g^V$ for the different species of
leptons, hadrons and quarks are given in the table below.

\begin{center}
\begin{tabular}{|c|c| }
\hline
~Particles ~&~ $g^{V}$ \\
\hline
~\textrm{$\nu_e,\nu_\mu,\nu_\tau$}~ & ~$\frac12$ \\
\hline
~\textrm{$e, \mu, \tau$}~ & ~$-\frac12+2\sin^2\theta_w$ \\
\hline
~\textrm{$p$}~ & ~$\frac12-2\sin^2\theta_w$ \\
\hline
~\textrm{$n$}~ & ~$-\frac12$ \\
\hline
~\textrm{$u, c, t$}~ & ~$\frac12-\frac{4}{3}\sin^2\theta_w$\\
\hline
~\textrm{$d, s, b$}~ & ~$-\frac12+\frac{2}{3}\sin^2\theta_w$\\
\hline
\end{tabular}\label{assign}
\end{center}

 \no Writing $ \Sigma^{f\,tad}(\om,k) $ in the
same form as (\ref{ReSelf}), we obtain \be
\sigma^0_{tad}=\frac{2\,G_F\,g^V}{\sqrt2}\,(n_f-n_{\bar{f}}) \quad ,
\quad \sigma^1_{tad}=0 \; . \ee \no and tadpoles with quark loops
acquire an extra factor $3$ from color. The tadpole contribution is
  proportional to $n_f-n_{\bar{f}}$, which is the signature
of a $CP$ asymmetric medium.   It is customary and convenient  to
express $n_f-n_{\bar{f}}$ in terms of its relative abundance to the
photons in the universe. At any temperature $T$, the photon number
density in the universe is given by \be n_\ga=\frac{2}{\pi^2} \;
\zeta(3) \; T^3 \; . \ee \no and the particle-antiparticle asymmetry
for any fermion species $f$ is defined as \be
L_f=\frac{n_f-n_{\bar{f}}}{n_\ga} \; . \ee The magnitude of the
observed baryon asymmetry is $ L_B \simeq 10^{-9} $. Since $ B-L $
is conserved in the standard model, it is natural to expect that the
lepton asymmetries should be of the same order as $ L_B $. Although
there is no a priori reason to expect the neutrino asymmetries to be
of the same order, we will henceforth \emph{assume} that all the
lepton and neutrino asymmetries $L_e, \; L_\mu , \; L_{\nu_e} $ and
$ L_{\nu_{\mu}} $ are of the order of the baryon asymmetry $ \sim
10^{-9} $.

The next step is to compute the contributions from the one-loop
exchange diagrams. We focus on two different temperature regimes, $
m_e \ll T \ll m_\mu $ in which only the electron is
ultrarelativistic and $ m_e, \; m_\mu \ll T \ll M_W $ in which both
leptons are ultrarelativistic.

\subsubsection{   $\mathbf{m_e\ll T \ll m_\mu}$}

This temperature limit is interesting because it is the energy scale
right above BBN. The efficiency of BBN is sensitive to the amount of
electron neutrinos which, in turn, depends on the detailed dynamics
of neutrino oscillations.  As it will be discussed below in detail,
in the temperature regime $ T  \gtrsim 5\, \textrm{MeV} $ the
non-local  (in space-time) terms proportional to $\omega,k$ from the
exchange diagrams (both charged and neutral currents) dominate the
contributions of the lepton and neutrino asymmetries \emph{assuming}
all of them to be of order $ 10^{-9} $.

In the temperature limit with $ m_{e} \ll T \ll m_{\mu} $,  the
contribution from $\mu$ leptons to the exchange one loop diagram is
exponentially suppressed and we neglect it.     Apart from electrons
and neutrinos, the thermal background does contain protons and
neutrons in nuclear statistical equilibrium since for $ T \gtrsim
1\,\textrm{MeV} $ the weak interactions lead to equilibration on
time scales shorter than the Hubble time scale via the reactions \be
n\leftrightarrow p+e^{-}+\bar{\nu}_e \quad , \quad p+\bar{\nu}_e
\leftrightarrow n+e^{+} \quad , \quad p+e^{-}\leftrightarrow n+\nu_e
\; . \ee In the basis of flavor eigenstates, the total one-loop
self-energy contribution is of the form \be \label{ReSig}
\textit{Re}\,\Sigma(\om,k) =[\gamma^0 \mathds{A}(\om
)-\vec{\ga}\cdot\uvk \,\mathds{B}( k)]\,L \ee \noindent where
$\mathds{A}(\om,k)$ and $\mathds{B}(\om, k)$ are $2\times2$ diagonal
matrices in the neutrino flavor basis given by \bea \mathds{A}(\om
)= \left(
                   \begin{array}{cc}
                    A_e(\om ) & 0 \\
                    0& A_\mu(\om ) \\
                   \end{array}
                 \right)~;~~
\mathds{B}(k) = \left(
                   \begin{array}{cc}
                    B_e( k) & 0 \\
                    0& B_\mu( k) \\
                   \end{array}
                 \right),
\eea Extracting non-local terms  (in space-time) up to
$\mathcal{O}(\omega/T)\,,\,\mathcal{O}(k/T)$ we find the following
matrix elements, \bea A_e(\om )&=&-\frac{3 \; G_F}{\sqrt2}\,L_e \;
n_\ga+\frac{7 \; \pi^2}{15 \; \sqrt2} \; \frac{G_F \; \om \;
T^4}{M_W^2}-\frac{3 \; G_F}{2 \; \sqrt2} \; L_{\nu_e} \; n_\ga+
\frac{7 \; \pi^2}{30 \; \sqrt2} \; \frac{G_F \; \om \;
T^4}{M_Z^2}\nonumber \\
&&+ \frac{G_F}{\sqrt2}\; L_{\nu_e} \; n_\ga+\frac{G_F}{\sqrt2} \;
L_{\nu_\mu} \;  n_\ga +\frac{2 \; G_F}{\sqrt2}(-\frac12+2 \; \sin^2
\; \theta_w) \; L_e \;
n_\ga \nonumber \\
&&+\frac{2 \; G_F}{\sqrt2}(\frac12-2 \; \sin^2\theta_w) \; L_p \;
n_\ga-\frac{G_F}{\sqrt2} \; L_{n} \; n_\ga, \\
A_{\mu}(\om)&=&-\frac{3 \; G_F}{2 \; \sqrt2} \; L_{\nu_\mu} \;
n_\ga+
\frac{7 \; \pi^2}{30 \; \sqrt2} \; \frac{G_F \; \om \; T^4}{M_Z^2}\nonumber \\
&&+\frac{G_F}{\sqrt2} \; L_{\nu_e} \; n_\ga+\frac{G_F}{\sqrt2} \;
 L_{\nu_\mu} \;
n_\ga +\frac{2 \; G_F}{\sqrt2}(-\frac12+2 \; \sin^2\theta_w)\,L_e \;
n_\ga \nonumber \\
&&+\frac{2 \; G_F}{\sqrt2}(\frac12-2 \; \sin^2\theta_w) \; L_p \;
n_\ga-\frac{G_F}{\sqrt2} \; L_{n} \; n_\ga \; , \\
B_e( k)&=&-\frac{7 \; \pi^2}{45 \; \sqrt2} \; \frac{G_F \;  k \;
T^4}{M_W^2}\left[1-\frac{30}{7 \;
\pi^2}\left(\frac{m_e}{T}\right)^2\right]
-\frac{7 \; \pi^2}{90 \; \sqrt2} \; \frac{G_F \;  k \;  T^4}{M_Z^2} \; ,\\
B_\mu( k)&=&-\frac{7 \; \pi^2}{90 \; \sqrt2}\frac{G_F \;  k \;
T^4}{M_Z^2}. \eea We purposely displayed the individual terms in the
above expressions to highlight that the first line  in
$A_{e,\mu}(\omega )$ as well as the expressions for $B_{e,\mu}( k)$
arise from the exchange diagrams (the two top diagrams in
fig.\ref{neusigma}), while the second and third lines in
$A_{e,\mu}(\omega )$ arise from the tadpole diagrams (bottom two
diagrams in fig.\ref{neusigma}). We have assumed that the flavor
neutrinos are in thermal equilibrium and have consistently neglected
neutrino masses. The correction term $(m_e/T)^2$ is displayed so
that one can estimate the error incurred when this term is dropped
for $T>>m_e$, the error is less than $1\%$ for $T \gtrsim 5
\,\textrm{MeV}$. In what follows we will neglect this contribution
in the temperature range of interest for this section.

Charge neutrality requires that  $L_e=L_p$, hence \be \Sigma^{e
\,tad} + \Sigma^{p \,\, tad}=0 \; . \ee Therefore, the expressions
for $A_e$ and $A_\mu$ simplify to the following: \bea A_e(\om )&=&
\frac{G_F\;n_\ga}{\sqrt2}\Bigg[-\mathcal{L}_e +\frac{7 \; \pi^4}{60
\; \zeta(3)} \; \frac{\om \;
T}{M_W^2}\left(2+\cos^2\theta_w \right)\Bigg] \; , \label{Aelo} \\
A_\mu(\om )&=& \frac{G_F \; n_\ga}{\sqrt2}\Bigg[-\mathcal{L}_\mu
+\frac{7 \; \pi^4}{60 \; \zeta(3)} \; \frac{\om \;
T}{M_W^2}\cos^2\theta_w \Bigg] \; ,
\label{Aulo}\\
B_e( k)&=& -\frac{G_F\;n_\ga}{\sqrt2} \frac{7 \; \pi^4}{180 \;
\zeta(3)} \; \frac{k \;
T}{M_W^2}\left(2+ \cos^2\theta_w \right) \; ,\label{Belo} \\
B_\mu( k)&=& -\frac{G_F\;n_\ga}{\sqrt2}\frac{7 \; \pi^4}{180 \;
\zeta(3)} \; \frac{k \; T}{M_W^2}\cos^2\theta_w \;  . \label{Bulo}
\eea \noindent where \bea -\mathcal{L}_e  & = & -\frac12 \;
L_{\nu_e}+L_{\nu_{\mu}}-3 \; L_e-L_n
\nonumber \\
-\mathcal{L}_\mu  & = & -\frac12L_{\nu_{\mu}}+L_{\nu_e}-L_n
\label{els} \eea Where in the temperature regime $ m_e \ll T \ll
m_\mu $ we consistently  neglected the muon contribution to the
non-local   terms proportional to $ \omega,k $ in the exchange
diagrams.

The above expressions also reveal the importance of the temperature
region $ T \gtrsim 5 \,\textrm{MeV} $. \emph{Assuming} that all
asymmetries are of the same order as the baryon asymmetry, namely $
L_i \sim 10^{-9} $, we see that for $ \omega \sim k \sim T $ the
factor $ T^2/M^2_W \gg 10^{-9} $ for $ T \gtrsim 5 \,\textrm{MeV} $.

We will discuss below that in this region there is also a resonance
in the mixing angle in agreement with the results in\cite{barbieri}.

\subsubsection{$\mathbf{m_e, m_\mu \ll T \ll M_W}$}

This temperature region is important because the non-local
contributions are much larger than that of the lepton and neutrino
asymmetries, assuming \emph{both} to be of the same order $\sim
10^{-9}$ and are the \emph{same} for both leptons if their masses
are neglected. Therefore, if the contribution from the lepton and
neutrino asymmetries is neglected, \emph{and} terms of
$\mathcal{O}(m^2_e/T^2)\,;\,\mathcal{O}(m^2_\mu/T^2)$ are
neglected, the matrices $\mathds{A},\mathds{B}$ become
proportional to the identity. In this case the mixing angle would
be the same as in the vacuum. We will see however, that keeping
terms of $\mathcal{O}(m^2_e/T^2)\,;\,\mathcal{O}(m^2_\mu/T^2)$
leads to a very different result, namely the vanishing of the
mixing angle for negative helicity neutrinos or positive helicity
antineutrinos in this temperature range.

For $T \gg m_\mu \sim 100\,\textrm{MeV}$ the temperature is larger
than the critical temperature for deconfinement in  QCD  $T_c\sim
160$MeV. Therefore, the medium contains free  quarks but  no
nucleons. Since both $u$ and $d$ quarks have masses smaller than
$10$ MeV, their masses can be neglected. We only include in our
description the two lightest quark degrees of freedom consistently
with keeping only a weak doublet. We also assume that there is
vanishing  strangeness in the medium.

As a result, the corresponding $ A_{e,\mu}(\om ), B_{e,\mu}( k) $
  are now given by
\bea A_e(\om )&=&-\frac{3 \; G_F}{\sqrt2} \; L_e \; n_\ga+\frac{7 \;
\pi^2}{15 \; \sqrt2}\frac{G_F \; \om \; T^4}{M_W^2}-\frac{3 \;
G_F}{2\sqrt2} \; L_{\nu_e} \; n_\ga+
\frac{7 \; \pi^2}{30 \; \sqrt2}\frac{G_F \; \om \; T^4}{M_Z^2}\nonumber \\
&&+ \frac{G_F}{\sqrt2} \; L_{\nu_e} \; n_\ga+\frac{G_F}{\sqrt2} \;
L_{\nu_\mu} \; n_\ga \nonumber \\
&&+\frac{2 \; G_F}{\sqrt2}(-\frac12+2 \; \sin^2\theta_w) \; L_e \;
n_\ga+\frac{2 \; G_F}{\sqrt2}(-\frac12+2 \; \sin^2\theta_w) \; L_\mu
\;
n_\ga \nonumber \\
&&+\frac{6 \; G_F}{\sqrt2}(\frac12-\frac43 \; \sin^2\theta_w) \; L_u
\; n_\ga+\frac{6 \; G_F}{\sqrt2}(-\frac12+\frac23 \; \sin^2\theta_w)
\; L_d \;
n_\ga \; , \\
A_{\mu}(\om )&=&-\frac{3 \; G_F}{\sqrt2} \; L_\mu \; n_\ga+\frac{7
\; \pi^2}{15 \; \sqrt2}\frac{G_F \; \om \; T^4}{M_W^2}-\frac{3 \;
G_F}{2 \; \sqrt2} \; L_{\nu_\mu} \; n_\ga+
\frac{7 \; \pi^2}{30 \; \sqrt2}\frac{G_F \; \om \; T^4}{M_Z^2}\nonumber \\
&&+\frac{G_F}{\sqrt2} \; L_{\nu_e} \; n_\ga+\frac{G_F}{\sqrt2}
\; L_{\nu_\mu} \; n_\ga \nonumber \\
&&+\frac{2 \; G_F}{\sqrt2}(-\frac12+2 \; \sin^2\theta_w) \; L_e \;
n_\ga+\frac{2 \; G_F}{\sqrt2}(-\frac12+2 \; \sin^2\theta_w) \; L_\mu
\;
n_\ga \nonumber \\
&&+\frac{6 \; G_F}{\sqrt2}(\frac12-\frac43 \; \sin^2\theta_w) \; L_u
\; n_\ga+\frac{6 \; G_F}{\sqrt2}(-\frac12+\frac23 \; \sin^2\theta_w)
\; L_d \;
n_\ga \; , \\
B_e( k)&=&-\frac{7 \; \pi^2}{45 \; \sqrt2}\frac{G_F \;  k \;
T^4}{M_W^2}\left[1-\frac{30}{7 \;
\pi^2}\left(\frac{m_e}{T}\right)^2\right]
-\frac{7 \; \pi^2}{90 \; \sqrt2}\frac{G_F \;  k \;  T^4}{M_Z^2} \; ,\\
B_\mu( k)&=&-\frac{7 \; \pi^2}{45 \; \sqrt2}\frac{G_F \;  k \;
T^4}{M_W^2}\left[1-\frac{30}{7 \;
\pi^2}\left(\frac{m_\mu}{T}\right)^2\right] -\frac{7 \; \pi^2}{90 \;
\sqrt2}\frac{G_F \;  k \;  T^4}{M_Z^2} \; . \eea Charge neutrality
of the medium leads to the constraint $ 4 \; L_u-L_d-3 \; L_e=0 $,
which leads to the following simplified expressions: \bea A_e(\om
)&=& \frac{G_F \; n_\ga}{\sqrt2}\Bigg[\ \widetilde{-\mathcal{L}_e}
+\frac{7 \; \pi^4}{60 \; \zeta(3)} \; \frac{\om \;
T}{M_W^2}\left(2+ \cos^2\theta_w \right) \Bigg]\label{Aehi} \; , \\
 A_\mu(\om )&=&
\frac{G_F \; n_\ga}{\sqrt2}\Bigg[-\widetilde{\mathcal{L}_\mu} +
\frac{7 \; \pi^4}{60 \; \zeta(3)} \; \frac{\om \;
T}{M_W^2}\left(2+ \cos^2\theta_w \right)\Bigg]\label{Auhi} \; , \\
B_e( k)&=& -\frac{G_F \; n_\ga}{\sqrt2} \frac{7 \; \pi^4}{180 \;
\zeta(3)}\frac{k \;  T}{M_W^2}\left[2+ \cos^2\theta_w - \frac{60}{7
\; \pi^2}\left(\frac{m_e}{T}\right)^2\right]
\label{Behi} \; ,\\
B_\mu( k)&=& -\frac{G_F \; n_\ga}{\sqrt2} \frac{7 \; \pi^4}{180 \;
\zeta(3)}\frac{k \;  T}{M_W^2}\left[2+ \cos^2\theta_w -
\frac{60}{7\pi^2}\left(\frac{m_\mu}{T}\right)^2\right] \; ,
\label{Buhi} \eea \noindent where \bea -\widetilde{\mathcal{L}_e} &
= & -\frac12 \; L_{\nu_e}+L_{\nu_{\mu}}-3 \; L_e + ( 1-4 \;
\sin^2\theta_w)(2 \; L_e-L_\mu)-
(1-8\sin^2\theta_w)L_u-2 \; L_d   \; ,\\
-\widetilde{\mathcal{L}_\mu}  & = & -\frac12 \;
L_{\nu_{\mu}}+L_{\nu_e}-3 \; L_\mu + ( 1-4 \; \sin^2\theta_w)(2 \;
L_e-L_\mu)-(1-8 \; \sin^2\theta_w)L_u-2 \; L_d \label{widels} \; .
\eea In the limit when $T\gg m_{e,\mu}$ both leptons become
ultrarelativistic and a CP-symmetric medium becomes flavor blind to
the weak interactions. In this case we must keep terms of $
\mathcal{O}(m_{e,\mu}/T) $ to understand the nature of oscillations
and mixing.

\section{Dispersion relations,
mixing angles and resonances in the medium}\label{sec:diagonal}

The neutrino dispersion relations and mixing angles in the medium
are obtained by diagonalizing the homogeneous effective Dirac
equation in the medium, namely by setting $ \eta(\omega,k)=0 $ in
eq.(\ref{EqMotFT}). Using the results obtained above, the
homogeneous Dirac equation in frequency and momentum becomes \be
\label{DiracEq} \left[\gamma^0\,\omega\mathds{1}- \vec{\gamma}\cdot
\uvk\, k \mathds{1}-\mathds{M}+ \left(\gamma^0 \mathds{A}(\om
)-\vec{\ga}\cdot\uvk \,\mathds{B}( k) \right)L \right]\,\psi
(\omega,k)= 0 \; , \ee \no where $\mathds{1}$ is the $2\times2$
identity matrix in the flavor basis in which the field $\psi(\om,k)$
is given by \be \psi(\om,k)=\left(
                  \begin{array}{c}
                    \nu_e(\om,k) \\
                    \nu_{\mu}(\om,k) \\
                  \end{array}
                \right) \; ,
\ee \no with $\nu_e(\om,k)$ and $\nu_{\mu}(\om,k)$ each being a
4-component Dirac spinor.

If we multiply the effective Dirac equation (\ref{DiracEq}) by the
chiral projectors $R$ and $L$ respectively from the left, we obtain
\bea &&\left(\gamma^0\,\mathds{W} - \vec{\gamma}\cdot \uvk\,
\mathds{K} \right)\psi_L-\mathds{M} \; \psi_R =0 \;  ,\label{EqL}\\
&&\left(\gamma^0\,\omega \,\mathds{1}- \vec{\gamma}\cdot \uvk\, k
\,\mathds{1} \right)\psi_R-\mathds{M} \; \psi_L =0 \label{EqR} \;
, \eea \no where we have defined the flavor matrices \be  {
\mathds{W}}=\om \mathds{1}+\mathds{A} \quad , \quad {\mathds{K}}=k
\, \mathds{1}+\mathds{B} \; . \ee The set of equations (\ref{EqL})
and (\ref{EqR}) couple $\psi_L$ and $\psi_R$ together. To solve
the equations, we first multiply (\ref{EqL}) by
$\left(\gamma^0\,\omega- \vec{\gamma}\cdot \uvk\,
k\right)\,\mathds{1}$ from the left and use eq.(\ref{EqR}) to
obtain an equation for $\psi_L$ which can be written in terms of
the helicity operator  $ \hat{h}(\uvk)=\gamma^0 \vec{\gamma}\cdot
\uvk\, \ga^5,  $ as follows \be \left[\om\,{ \mathds{W}}-k\,{
\mathds{K}}+\hat{h}(\uvk) (\om { \mathds{K}} - { \mathds{W}}
k)-\mathds{M}^2\right]\psi_L=0 \; , \label{psiLeq} \ee and the
right handed component is given by \be \label{eignEqR}
\psi_R(\omega,k) = \mathds{M}\,\gamma^0\, \frac{[\om
+\hat{h}(\uvk) \; k]}{\omega^2-k^2} \; \psi_L(\omega,k) \; . \ee
It is convenient to separate the Dirac and flavor structure to
simplify the study. This is achieved most economically in the
chiral representation of the Dirac matrices, in which \bea
\ga^0=\left(
             \begin{array}{cc}
               0 & \mathds{-1} \\
               \mathds{-1} & 0 \\
             \end{array}
           \right)~;~~\ga^5=\left(
                        \begin{array}{cc}
                          \mathds{1} & 0 \\
                          0 & \mathds{-1} \\
                        \end{array}
                      \right) \; ,
\eea \bea \vec{\ga}\cdot\uvk=\left(
             \begin{array}{cc}
               0 & \vec{\sigma}\cdot\uvk \,\mathds{1} \\
              -\vec{\sigma}\cdot\uvk \,\mathds{1}  & 0 \\
             \end{array}
           \right)~;~~\hat{h}(k)=\vec{\sigma}\cdot\uvk \left(
                        \begin{array}{cc}
                          \mathds{1} & 0 \\
                          0 & \mathds{1} \\
                        \end{array}
                      \right) \; ,
\eea \noindent and by introducing the two component Weyl spinors
$v^{(h)}(\uvk)$ eigenstates of helicity, \be \label{weyl}
\vec{\sigma}\cdot\uvk \,v^{(h)}(\uvk) = h \, v^{(h)}(\uvk) ~~;~~
h=\pm 1 \; . \ee These spinors are normalized so that \be
\left(v^{(h)}(\uvk)\right)^\dagger v^{(h')}(\uvk) = \delta_{h,h'}
\label{normaW}  \; . \ee In terms of these helicity eigenstates, a
general flavor doublet of  left (L) and right (R)  handed Dirac
spinors can be written \be \psi_L = \sum_{h=\pm1} \left(
                        \begin{array}{ c}
                            0 \\
                          v^{(h)}\otimes \varphi^{(h)} \\
                        \end{array}
                      \right) \label{geneL} \; ,
\ee \noindent and \be \psi_R = \sum_{h=\pm1} \left(
                        \begin{array}{ c}
                            v^{(h)}\otimes \xi^{(h)} \\
                          0 \\
                        \end{array}
                      \right) \label{geneR} \; ,
\ee
 \noindent where $ \varphi^{(h)}~;~\xi^{(h)} $ are flavor
doublets. We have purposely left the arguments unspecified because
this expansion will be used in real time as well as for the Fourier
and Laplace transforms respectively. We need both positive and
negative helicity eigenstates because the four independent degrees
of freedom for each flavor are positive and negative energy and
positive and negative helicity.

Projecting eq. (\ref{psiLeq}) onto the helicity eigenstates
$v^{(h)}(\uvk)$ we obtain an equation for the flavor doublet
$\varphi^{(h)} (\omega,k)$, namely \be \label{eignEqL}
\left[(\om^2-k^2)\mathds{1}+(\om-h k)(\mathds{A}+h \mathds{B})-
\mathds{M}^2 \right]\varphi^{(h)} (\omega,k)=0 \; . \ee Projecting
eq. (\ref{eignEqR}) onto helicity eigenstates yields the relation
\be \label{xieq} \xi^{(h)}(\omega,k) = -\mathds{M}  \; \frac{(\om +h
\;  k) }{\omega^2-k^2} \; \varphi^{(h)}(\omega,k) \; . \ee Writing
the doublet $\varphi^{(h)}(\omega,k)$ in the flavor basis as \be
\varphi^{(h)} (\omega,k) = \left(
         \begin{array}{c}
           \nu^{(h)}_{e } (\omega,k)\\
           \nu^{(h)}_{\mu  }(\omega,k) \\
         \end{array}
       \right) \label{flavdou} \; ,
\ee \noindent  leads to the following matrix form for
Eq.(\ref{eignEqL}) \bea \label{matrixEq} \left(
  \begin{array}{cc}
    a^{(h)} & b \\
    b & c^{(h)} \\
  \end{array}
\right)\left(
         \begin{array}{c}
           \nu^{(h)}_{e } \\
           \nu^{(h)}_{\mu  } \\
         \end{array}
       \right)=0 \; ,
\eea \no where the matrix elements in the flavor basis are given by
\bea a^{(h)}&=&\om^2-k^2+(\om-h k)(A_e+h B_e)-\frac12 \;
(M_1^2+M_2^2) -\frac12 \; (M_1^2-M_2^2) \; \cos(2\theta)
\; ,\\ b&=& \frac12 \; (M_1^2-M_2^2)\sin(2\theta) \; , \\
c^{(h)}&=&\om^2-k^2+(\om-h k)(A_\mu+h B_{\mu})-\frac12 \;
(M_1^2+M_2^2) +\frac12 \; (M_1^2-M_2^2) \; \cos(2\theta) \; . \eea
Let us introduce a doublet of mass eigenstates in the medium \be
\label{massei} \chi^{(h)}(\om,k) = \left(
                  \begin{array}{c}
                    \nu_1(\om,k) \\
                    \nu_2(\om,k) \\
                  \end{array}
                \right) \; ,
\ee \noindent related to the flavor doublet  $
\varphi^{(h)}(\omega,k) $ by a unitary transformation   $ U^{(h)}_m
$ with \be \label{mediumrotmat} U^{(h)}_m = \left(
          \begin{array}{cc}
            \cos\theta_m^{(h)} & \sin \theta_m^{(h)} \\
            -\sin \theta_m^{(h)} & \cos \theta_m^{(h)} \\
          \end{array}
        \right) \; ,
\ee \be  \label{unitrafo2} \varphi^{(h)}(\omega,k) = U^{(h)}_m \;
\chi^{(h)}(\omega,k)~~;~~  \xi^{(h)}(\omega,k) = U^{(h)}_m \;
\zeta^{(h)}(\omega,k) \; . \ee The mixing angle in the medium for
states with helicity $ h, \; \theta_m^{(h)} $ is obtained by
requiring that the unitary transformation eq.(\ref{mediumrotmat})
diagonalizes the matrix equation eq.(\ref{matrixEq}). The eigenvalue
equation in diagonal form is given by \be  \Bigg\{ \omega^2-k^2
+\frac12 \;  S_h(\omega,k)-\frac12 \; (M^2_1+M^2_2)-\frac12 \;
\delta\,M^2 \left[\left(\cos 2\theta-\frac{\Delta_h
(\omega,k)}{\delta M^2}\right)^2+\sin^22\theta
\right]^{\frac12}\left(
          \begin{array}{cc}
           1 & 0 \\
            0 & -1 \\
          \end{array}
        \right)\Bigg\}\chi^{(h)}(\omega,k)
          =0 \label{psimass}  \; ,
\ee \noindent where $ S_h(\omega,k) ,  \;  \delta M^2 $ and $
\Delta_h $ are respectively given by \bea S_h(\omega,k) & = & (\om-h
k)\left[A_e(\omega )+A_\mu(\omega ) +h  \; B_e( k)+h \; B_\mu(
k)\right]
\label{Sofome} \; , \\ \delta M^2 & = & M_1^2-M_2^2\label{delM2}  \; , \\
\Delta_h(\omega,k) &  = & (\om-h k)\left[A_e(\omega )-A_\mu(\omega
)+h \; B_e( k)-h \; B_\mu( k) \right]\label{Deltah} \; . \eea The
mixing angle in the medium is determined by  the relation \be
\label{tanmix} \tan [2\theta_m^{(h)}] = \frac{2 \;
b}{c^{(h)}-a^{(h)}} = \frac{\delta M^2 \; \sin(2\theta)}{\delta M^2
\;  \cos(2\theta)-\Delta_h(\omega,k)} \; , \ee or alternatively by
the more familiar relation \be \label{sinmix}\sin2\theta^{(h)}_m =
\frac{\sin2\theta}{\left[\left(\cos
2\theta-\frac{\Delta_h(\omega,k)}{\delta M^2}\right)^2+\sin^22\theta
\right]^{\frac12}} \ee We note that the neutrino mass eigenvalues as
well as the mixing angle depends on $  k $ as well as on the \emph{
helicity eigenvalue} $ h $. This is one of the novel results which
has not been obtained before simply because only left handed
negative helicity neutrinos were considered in the literature
\cite{MSWI,MSWII,bethe,grimus,gouvea,book1}.

The right handed components are obtained from the left handed ones
by performing the unitary transformation eq.(\ref{mediumrotmat}) on
eq.(\ref{xieq}). The relation (\ref{xieq}) leads to the following
expressions \be \label{rightrel}
\zeta^{(h)}(\omega,k)=-\frac{\omega+h\,k}{\omega^2-k^2} \;
\overline{M}  \; \Bigg[ \mathds{1}+ \frac{\delta
M^2}{4{\overline{M}}^2}\left(
          \begin{array}{cc}
            \overline{C} &\overline{S} \\
            \overline{S}& -\overline{C} \\
          \end{array}
        \right) \Bigg]\chi^{(h)}(\omega,k) \; ,
 \ee
\noindent where $ \overline{M} =\frac12(M_1+M_2) $ and

\be  \overline{C}= \cos\left[2 \; \theta^{(h)}_m -2 \; \theta
\right] \quad , \quad \overline{S}= \sin\left[2 \; \theta^{(h)}_m -2
\; \theta \right]  \; . \ee
\subsection{Eigenvectors and dispersion relations}

Eq.(\ref{psimass}) has the following eigenvectors in the basis of
mass eigenstates: \bea \chi^{(h)}_{1 }(\omega,k)  &= &  \nu^{(h)}_{
1}(\omega,k)\, \left(
          \begin{array}{c}
            1 \\
            0 \\
          \end{array}\right) \label{chi1} \; , \\ \zeta^{(h)}_1(\omega,k) & = &
-\nu^{(h)}_{  1}(\omega,k) \; \frac{\omega+h\,k}{\omega^2-k^2} \;
\overline{M} \;  \Bigg[ \left(
          \begin{array}{c}
            1 \\
            0 \\
          \end{array}\right)+
\frac{\delta M^2}{4 \; {\overline{M}}^2}\left(
          \begin{array}{c}
            \overline{C} \\
           \overline{S}\\
          \end{array}\right)\Bigg] \label{zeta1}  \; ,
\eea
 \noindent and
\bea
          \chi^{(h)}_{2 }(\omega,k)  & = & \nu^{(h)}_{2}(\omega,k)\,\left(
          \begin{array}{c}
            0 \\
            1 \\
          \end{array}\right) \; ,\label{chi2}\\
\zeta^{(h)}_2(\omega,k) & = & -\nu^{(h)}_{  2}(\omega,k) \;
\frac{\omega+h\,k}{\omega^2-k^2} \; \overline{M} \;  \Bigg[ \left(
          \begin{array}{c}
            0 \\
           1 \\
          \end{array}\right)+
\frac{\delta M^2}{4 \; {\overline{M}}^2}\left(
          \begin{array}{c}
            \overline{S} \\
           -\overline{C}\\
          \end{array}\right)\Bigg] \label{zeta2}
\eea The corresponding doublets in the flavor basis can be obtained
by   the unitary transformation  eq.(\ref{mediumrotmat}).

The eigenvalues are found in perturbation theory consistently up
to $\mathcal{O}(G_F)$ by writing \be  \omega^{(h)}_a(k,\pm)   =
\pm \left[E_a(k)+\delta\omega^{(h)}_a(k,\pm)\right] \quad , \quad
a=1,2 \label{omegaa} \ee with \be E_{1,2}(k) =
\sqrt{k^2+M^2_{1,2}} \ee We find, \be \delta\omega^{(h)}_1(k,\pm)
= -\frac{1}{4E_1(k)}\Bigg\{S_h(\pm E_1(k),k)-\delta M^2
\Bigg[\left[\left(\cos 2\theta-\frac{\Delta_h (\pm
E_1(k),k)}{\delta M^2}\right)^2+\sin^22\theta
\right]^{\frac12}-1\Bigg] \Bigg\} \label{delom1} \ee \be
\delta\omega^{(h)}_2(k,\pm) = -\frac{1}{4E_2(k)}\Bigg\{S_h(\pm
E_2(k),k)+\delta M^2 \Bigg[\left[\left(\cos 2\theta-\frac{\Delta_h
(\pm E_2(k),k)}{\delta M^2}\right)^2+\sin^22\theta
\right]^{\frac12}-1\Bigg] \Bigg\} \label{delom2} \ee It is
important to highlight that whereas the mixing angle only depends
on $ \Delta_h $, the medium corrections to the frequencies also
depend on $ S_h $. This is important because even when in the case
when   the matrices $ \mathds{A}, \; \mathds{B} $ become
proportional to the identity, in which case $\Delta_h=0$ and the
mixing angle in the medium coincides with that of the vacuum, the
frequencies and in particular the oscillation frequency   still
receives medium corrections.

\subsection{Resonances}\label{resonance}

The condition for resonant oscillations is that the mixing $ \tan
[2\theta_m^{(h)}] $ reaches a maximum (infinity) as a function of
a parameter, temperature, density or energy. From eq.
(\ref{tanmix}) a resonance takes place when \be \label{reso}
\frac{\Delta_h(\omega,k)}{\delta M^2} = \cos 2\theta \ee \noindent
where $ \omega= \omega^{(h)}_a(k,\pm) $   correspond to the
dispersion relations for the propagating modes in the medium,
given by eq.(\ref{omegaa}). To leading order in $ G_F $ the
in-medium dispersion relation can be approximated by the free
field dispersion relation $ \omega^{(h)}_a(k,\pm)\approx \pm
\sqrt{k^2+M^2_{1,2}} $. The relativistic limit is warranted
because the neutrino momenta in the plasma is  $ k \gg M_{1,2}
\sim \mathrm{eV} $. Furthermore, under the assumption that the
hierarchy of vacuum mass eigenstates is nearly degenerate, namely
$ | \delta M^2/\overline{M}^{\,2} | \ll 1 $, as seems to be
supported by the experimental data, the dispersion relations can
be further approximated as follows
 \be
\omega^{(h)}_a(k,\pm)\approx   \lambda k\left(1 +
\frac{\overline{M}^{\,2}}{2k^2} \right)~~;~~ \lambda =\pm 1
\label{rela} \ee It is convenient  to introduce the following
notation \bea \mathcal{L}_9 & = & 10^9
\left(\mathcal{L}_e-\mathcal{L}_\mu\right)
\label{L9} \\
\delta_5 & = & 10^5 \left( \frac{\delta M^2}{\mathrm{eV}^2}\right)
\label{del5} \eea \emph{If} the lepton and neutrino asymmetries are
of the same order of the baryon asymmetry, then $ 0.2 \lesssim
|\mathcal{L}_9| \lesssim 0.7 $ and the fitting from solar and
KamLAND data suggests $|\delta_5 | \approx 8$. Using the
approximations leading to eq. (\ref{rela})  the ratio
$\Delta_h/\delta M^2$ can be written compactly from
eq.(\ref{Deltah}).

We study separately the cases $ m_e \ll T \ll m_\mu $ and $
m_e,m_\mu \ll T \ll M_W $.

\subsubsection{$ \mathbf{m_e \ll T \ll m_\mu}$}

\begin{itemize}
\item{ {\bf case I}:  $ \omega = k +
\frac{\overline{M}^{\,2}}{2k}~~,~~h=-1 $, \textit{positive energy,
negative helicity neutrinos}: \be \label{case1I}
\frac{\Delta_h}{\delta M^2} \approx
\frac{4}{\delta_5}\left(\frac{0.1\,T}{\mathrm{MeV}} \right)^4 \;
\frac{k}{T} \;  \Bigg[-\mathcal{L}_9 +
\left(\frac{2\,T}{\mathrm{MeV}} \right)^2\; \frac{k}{T} \Bigg] \;
. \ee Where we have neglected  $ \frac{\overline{M}^{\,2}}{k^2} $.
For fixed temperature, the resonance condition eq.(\ref{reso}) is
fulfilled for the value of neutrino momentum given by \be
\label{kresoI} k = \frac{(\mathrm{MeV})^2}{8
T}\Bigg\{\mathcal{L}_9 + \Bigg[\mathcal{L}^2_9 +
16\,\delta_5\,\cos 2\theta \left(\frac{200\,\mathrm{MeV}}{T}
\right)^2\Bigg]^{\frac12}\Bigg\} \; . \ee Hence, for $ \delta_5
\,\cos 2\theta > 0 $, there is always a resonance. \emph{If} $
|\mathcal{L}_9| \lesssim 1 $, then for neutrino momenta such that
$ \sqrt{T k}
> 1 \,\mathrm{MeV} $ the non-local term dominates over the asymmetry
and the resonance occurs for $ k \sim 25 \; \sqrt{\delta_5\;\cos
2\theta}\;(\mathrm{MeV})^3/T^2 $. For example, if $ T \sim 10\;
\mathrm{MeV} $, the resonance occurs for $ k\sim 1\; \mathrm{MeV} $.
If $ \delta_5 \,\cos 2\theta < 0 $ there can also be a resonance
provided \be \frac{|\mathcal{L}_9| \,T}{200\,\mathrm{MeV}}
> 4\sqrt{|\delta_5 \,\cos 2\theta|}. \ee}
\no However this inequality requires   a large value of
$|\mathcal{L}_9|$, for example for $T\sim 10 \mathrm{MeV}$ it
requires that  $|\mathcal{L}_9|\gtrsim 140 $. \item{ {{\bf case
II}: $\omega = k + \frac{\overline{M}^{\,2}}{2k}~~,~~h= 1$,
\textit{positive energy, positive helicity neutrinos:} } \be
\label{caseII1} \frac{\Delta_h}{\delta M^2} \approx
\frac{10^{-16}}{ \delta_5}\left(\frac{T}{\mathrm{MeV}} \right)^2\;
\left(\frac{\overline{M}}{\textrm{ eV} } \right)^2
\Bigg[-\mathcal{L}_9\; \frac{T}{k} + 2\left(\frac{
T}{\mathrm{MeV}} \right)^2 \Bigg] \ee Where we have neglected
terms of higher order in  $ \frac{\overline{M}^{\,2}}{k^2} $.
Because $ \overline{M} \sim 1\,\mathrm{eV} $  and $
100\,\mathrm{MeV}\gg T \gg 1 \, \mathrm{MeV} $ a resonance would
only be available for $ k \sim 10^{-16} \,\mathrm{MeV} $ which is
not a relevant range of momenta for neutrinos in the plasma.
Therefore, positive helicity neutrinos mix with the \emph{vacuum
mixing angle}.  } \item{ {{\bf case III}: $\omega = -k -
\frac{\overline{M}^{\,2}}{2k}~~,~~h=-1$, \textit{positive energy,
negative helicity anti-neutrinos:} } \be \frac{\Delta_h}{\delta
M^2} \approx \frac{10^{-16}}{ \delta_5}\left(\frac{
T}{\mathrm{MeV}} \right)^2\;\left(\frac{\overline{M}}{\textrm{ eV}
} \right)^2\Bigg[ \mathcal{L}_9\; \frac{T}{k} + 2\left(\frac{
T}{\mathrm{MeV}} \right)^2 \Bigg]  \; . \ee Again in this
expression we have neglected higher order terms in $
\frac{\overline{M}^{\,2}}{k^2} $. A conclusion similar to that of
case II above holds in this case. No resonance is available for
relevant values of neutrino momenta within the temperature range
in which these results are valid. For the cases II and III the
ratio $ |\Delta_h/\delta M^2| \ll 1 $ for all relevant values of
the neutrino momentum within the temperature range in which these
results are valid. Therefore, negative helicity
\emph{antineutrinos} mix with the vacuum mixing angle, just as
positive helicity neutrinos. }

\item{ {{\bf case IV}:  $ \omega = -k -
 \frac{\overline{M}^{\,2}}{2k}~~,~~h= 1 $,
\textit{positive energy, positive helicity anti-neutrinos:} } \be
\frac{\Delta_h}{\delta M^2} \approx
\frac{4}{\delta_5}\left(\frac{0.1\,T}{\mathrm{MeV}} \right)^4\;
\frac{k}{T} \Bigg[ \mathcal{L}_9 + \left(\frac{2\,T}{\mathrm{MeV}}
\right)^2 \; \frac{k}{T} \Bigg] \; . \ee Where we have neglected
higher order terms in $ \frac{\overline{M}^{\,2}}{k^2} $. The
position of the resonance in this case is obtained from that in
case I above by the replacement $\mathcal{L}_9 \rightarrow
-\mathcal{L}_9$, namely for fixed temperature the resonance
condition is fulfilled at the value of $k$ given by \be
\label{kresoIV} k = \frac{(\mathrm{MeV})^2}{8
T}\Bigg\{-\mathcal{L}_9 + \Bigg[\mathcal{L}^2_9 +16\,
\delta_5\,\cos 2\theta \left(\frac{200\,\mathrm{MeV}}{T}
\right)^2\Bigg]^{\frac12}\Bigg\}\; . \ee Again in the temperature
range $ 1\,\mathrm{MeV} \ll T \ll 100\,\mathrm{MeV} $ there is a
resonance if $ \delta_5\,\cos 2\theta >0 $ (assuming that $
|\delta_5 \,\cos 2\theta| \sim 1 $).

Just as in case I, \emph{if} $ |\mathcal{L}_9| \lesssim 1 $ the
non-local term dominates over the asymmetry contribution for $
\sqrt{T k}\gtrsim 1 \,\mathrm{MeV} $  and the resonance occurs for $
k \sim 25 \sqrt{\delta_5\,\cos 2\theta}\,(\mathrm{MeV})^3/T^2 $. }

Cases III and IV reveal an interesting feature: \emph{only} the
asymmetry contribution changes sign between neutrinos and
antineutrinos whereas the non-local  (in space-time) term
\emph{remains the same}.

Together these expressions confirm that \emph{if} the lepton and
neutrino asymmetries are of the same order as the baryon asymmetry,
namely $ 0.2 \lesssim |\mathcal{L}_9| \lesssim 0.7 $, then the
non-local terms from the exchange diagrams dominate the self-energy
for $ T \gtrsim 3-5 \, \mathrm{MeV}$ unless the neutrino in the
plasma has a momentum $ k $ such that $ \sqrt{kT} \ll 0.5 \,
\mathrm{MeV} $.

\end{itemize}

In summary, for $m_e \ll T \ll m_\mu$ resonances occur in cases I
and IV when $ h \; \lambda < 0 $ ($ \lambda $ is the sign  of the
energy eigenvalue). For $ h \; \lambda
> 0 $ (cases II and III) no resonance is available for  neutrino momenta $ k
\sim T \sim \mathrm{few}\,\mathrm{MeV}$  and mixing angle in the
medium coincides with the vacuum value.
\subsubsection{$\mathbf{ m_e,m_\mu \ll T \ll M_W}$}

We use $m_\mu \approx 106\,\mathrm{MeV} $ and find the following
simple expressions

\begin{itemize}
\item{ {{\bf case I}:  $\omega = k +
\frac{\overline{M}^{\,2}}{2k}~~,~~h=-1$, \textit{positive energy,
negative helicity neutrinos}: } \be
\label{caseI2}\frac{\Delta_h}{\delta M^2} \approx \frac{0.4\times
10^{12}}{\delta_5}\left(\frac{ T}{\mathrm{GeV}} \right)^4\;
\frac{k}{T} \; \Bigg[4.83 \; \frac{k}{T} +10^{-3}\mathcal{L}_9
\Bigg] \ee In this case no resonance is available but for neutrinos
with extremely low energy and not relevant for the plasma. For
example for $T \sim \mathrm{GeV}$ only neutrinos with energy of a
few $\mathrm{eV}$ would be potentially resonant, but this momentum
range is not a relevant one for neutrinos in the plasma. For
neutrinos with energy larger than a few $keV$ the mixing angle
effectively vanishes. Therefore, we conclude that in this
temperature regime the mixing angle in the medium for negative
helicity neutrinos vanishes. }

\item{ {{\bf case II}: $\omega = k +
\frac{\overline{M}^{\,2}}{2k}~~,~~h= 1$, \textit{positive energy,
positive helicity neutrinos}: } \be \label{caseII2}
\frac{\Delta_h}{\delta M^2} \approx - \frac{
10^{-7}}{\delta_5}\left(\frac{ T}{\mathrm{GeV}} \right)^2 \;
\frac{T}{k} \; \left( \frac{\overline{M}}{\mathrm{eV}}\right)^2
\Bigg[4.83 \; \frac{k}{T} - 10^{-3}\mathcal{L}_9 \Bigg] \ee It is
clear that for the relevant regime of neutrino momenta in the plasma
$|\Delta_h/\delta M^2| \ll 1$. Hence the mixing angle in the medium
coincides with the vacuum mixing angle. Thus the conclusion in this
case is similar to that in the   case described by eq.
(\ref{caseII1}), namely positive helicity neutrinos undergo
oscillations in the medium with the \emph{vacuum} mixing angle. }

\item{ {{\bf case III}: $\omega = -k -
\frac{\overline{M}^{\,2}}{2k}~~,~~h=-1$, \textit{positive energy,
negative helicity anti-neutrinos}: } \be  \frac{\Delta_h}{\delta
M^2} \approx   -\frac{ 10^{-7}}{\delta_5}\left(\frac{
T}{\mathrm{GeV}} \right)^2\; \frac{T}{k} \left(
\frac{\overline{M}}{\mathrm{eV}}\right)^2 \Bigg[4.83 \;
\frac{k}{T}+10^{-3}\mathcal{L}_9 \Bigg] \ee The result in this case
is similar to that of case II above, negative helicity
\emph{antineutrinos} oscillate in the medium with the vacuum mixing
angle. }
\item{ {{\bf case IV}:  $\omega = -k -
\frac{\overline{M}^{\,2}}{2k}~~,~~h= 1$, \textit{positive energy,
positive helicity anti-neutrinos}: } \be \frac{\Delta_h}{\delta M^2}
\approx  \frac{0.4\times 10^{12}}{\delta_5}\; \left( \frac{
T}{\mathrm{GeV}} \right)^4\; \frac{k}{T} \Bigg[4.83\; \frac{k}{T}
-10^{-3}\mathcal{L}_9 \Bigg] \ee The conclusion in this case is
similar to that of the case described by eq. (\ref{caseI2}) above,
the mixing angle effectively vanishes and oscillations of positive
helicity \emph{antineutrinos} are suppressed in the medium in this
temperature range. }

\end{itemize}

Taken together the above analysis reveals that there is a resonance
in the oscillation of negative helicity neutrinos and positive
helicity antineutrinos (that is  $ h \; \lambda < 0 $) in the
temperature range $ m_e \ll T \ll m_{\mu} $ with a typical neutrino
momentum $ k \sim T \sim \mathrm{few}\,\mathrm{MeV} $. For $
m_e,m_{\mu}\ll T \ll M_W $ the mixing angle for negative helicity
neutrinos and positive helicity antineutrinos (that is  $ h \;
\lambda < 0 $) effectively vanishes in the medium, and in both
temperature ranges positive helicity neutrinos and negative helicity
antineutrinos undergo oscillations in the medium with the vacuum
mixing angle. We cannot yet conclude that positive helicity
neutrinos and negative helicity antineutrinos are sterile, before
studying the corrections to the oscillation frequencies.

\subsection{Oscillation frequencies and time scales}\label{osfreq}

The oscillation time scale   in the medium is given by \be
\tau_{m}^{(h)} (k,\la)=
\frac{1}{\left|\om_1^{(h)}(k,\la)-\om_2^{(h)}(k,\la)\right|} =
\frac{1}{\left|E_1(k)-E_2(k)+\delta \om_1^{(h)}(k,\la)-\delta
\om_2^{(h)}(k,\la) \right|} \; , \ee \no where
$E_{1,2}(k)=\sqrt{k^2+M_{1,2}^2}$ and   $\la=+1$ and $\la=-1$
correspond to neutrino and antineutrinos respectively.

The vacuum oscillation time scale   is \be  \label{vactau}\tau_{v}(k
) = \frac{1}{\left| \,E_1(k)-E_2(k) \right|} \; , \ee \noindent
therefore, in order to understand the loop corrections to the
oscillations time scales, it is convenient to study the ratio \be
\label{rattau}\frac{\tau_{v} (k )}{\tau_{m}^{(h)} (k,\la)} = \Bigg|
1+
 \frac{\delta \om_1^{(h)}(k,\la)-\delta
\om_2^{(h)}(k,\la)}{E_1(k)-E_2(k) } \Bigg|  \; . \ee
 Typical neutrino momenta in the plasma are ultrarelativistic, hence
 we approximate
\be E_1(k)-E_2(k)\approx\frac{ \delta M^2}{2k} = \frac{\delta_5}{2}
\frac{10^{-11}\,\mathrm{eV}}{(k/\mathrm{MeV})} \; . \ee \noindent
furthermore to leading order in $ G_F $  we replace $
\omega^{(h)}_a(k,\la)\approx \la k $ in the arguments of
$A_{e,\mu}(\omega)$.

The term $\delta \om_1^{(h)}(k,\la)-\delta \om_2^{(h)}(k,\la)$
represents the correction to the vacuum oscillation time scale due
to the medium effect. While the general form of these corrections
are cumbersome, we can extract simplified expressions in three
relevant limits.

\begin{itemize}

\item{{\bf I  Resonant case:}
$\frac{\Delta_h(\la E_{1,2},k)}{\delta M^2}\approx \cos2\theta$:\\

In section \ref{resonance} above  we found that resonant flavor
oscillations can occur only for $\la=+1, h=-1$ and $\la=-1, h=+1$.
In both these cases we obtain, \bea \label{resona} \delta
\om_1^{(h)}(k,\la)-\delta \om_2^{(h)}(k,\la)&=& -\frac{\delta M^2}{8
k^2}\Bigg[A_e(k)+A_\mu(k) +B_e(k)+B_\mu(k)\Bigg]  \nonumber \\ &&
-\frac{\delta M^2}{2 k^2}\left(1+\frac{\overline{M}^{\,2}}{2
k^2}\right)k \; (1-\sin 2\theta) \; . \eea }
\item{{\bf II vanishing mixing angle :}
$ \left|\frac{\Delta_h(\la E_{1,2},k)}{\delta M^2}\right|\gg 1 $:\\

In this limit, $\theta_m^{(h)} \sim 0$ and  neutrino flavor mixing
is suppressed. In section \ref{resonance} above, we found that this
occurs only for $\la=+1, h=-1$ or $\la=-1, h=+1$. Furthermore,
$\Delta_h(\la E_{1,2},k)$ is always positive definite in both
temperature limits considered here $m_e\ll T \ll m_\mu$ and
$m_e,m_\mu \ll T \ll M_W$.

In this case we obtain \bea \label{suppression} \delta
\om_1^{(h)}(k,\la)-\delta \om_2^{(h)}(k,\la)&=& -\frac{\delta M^2}{8
k^2}\Bigg[A_e(k)+A_\mu(k)
+B_e(k)+B_\mu(k)\Bigg]\nonumber \\
&&+ \,\text{sign}\,(\delta M^2)\Bigg[A_e(k)-A_\mu(k)
-B_e(k)+B_\mu(k)\Bigg] \; . \eea }
\item{{\bf III vacuum mixing:}
$\left|\frac{\Delta_h(\la E_{1,2},k)}{\delta M^2}\right|\ll 1$:\\

In this limit    $\theta^{(h)}_m \approx \theta$. In section
\ref{resonance} above we found that this case occurs for $\la=+1,
h=+1$ or  $\la=-1, h=-1$. In both these cases we find \bea
\label{vacuum} \delta \om_1^{(h)}(k,\la)-\delta
\om_2^{(h)}(k,\la)&=& -\frac{\delta M^2}{8 k^2}\Bigg[A_e(k)+A_\mu(k)
+B_e(k)+B_\mu(k)\Bigg]\nonumber \\
&&-\frac{\overline{M}^{\,2}}{4 k^2}\cos 2\theta
\,\Bigg[A_e(k)-A_\mu(k)+B_e(k)-B_\mu(k)\Bigg] \; . \eea }

\end{itemize}

 We now study these simplified expressions in the different regimes
 of temperature and for the different helicities components. The
 most relevant cosmological regime corresponds to momenta of the
 order of the temperature, hence we will focus on the regime in
 which the non-local  (in space-time) contributions from the exchange diagrams
 dominate over the lepton-neutrino asymmetries. Taken together these
 simplifications allow us to study the relevant cosmological range
 of neutrino energies in a clear manner.

\subsubsection{$\mathbf{m_e\ll T \ll m_\mu}$}

\begin{itemize}

\item{ {\bf case I}:  $\omega_{1,2}(k,\la) = \la \;  (k +
\frac{M_{1,2}^2}{2k})~~;~~\la =+1, h=-1$ and $ \la =-1, h=+1 $: \\

As observed in section (\ref{resonance}) the non-local terms
become dominant for  $\sqrt{T k}\geq 1 \mathrm{MeV}$ which is of
course consistent with the ultrarelativistic limit
$\overline{M}^{\,2}/2k^2 \ll 1$. Furthermore in the temperature
range of interest in this study, the factors $A_{e,\mu}(k)$ and
$B_{e,\mu}(k)$ are of the order $G_F k T^{4}/M_W^2\sim 10^{-9}
(T/\mathrm{GeV})^4\,k \ll k$. Therefore, near the  resonance which
occurs when ${T^2 k} \sim 25\,\sqrt{\delta_5 \cos 2\theta}
\,\mathrm{MeV}^3$, the expression (\ref{resona}) simplifies to
\bea \label{resosim} \delta \om_1^{(h)}(k,\la)-\delta
\om_2^{(h)}(k,\la)&\approx & -\frac{\delta M^2}{2 k} \; (1-\sin
2\theta) \; . \eea The ratio of the oscillation time scales
(\ref{rattau}) becomes \be \frac{\tau_v (k )}{\tau_m^{(h)}
(k,\la)} \sim  |\sin 2 \; \theta| <1 \ee Therefore, for small
\emph{vacuum} mixing angle there is a considerable \emph{slow
down} of oscillations. Resonant flavor mixing in the medium occurs
on \emph{longer} time scales than in the vacuum.

 For large neutrino energy, well outside the resonance region for
 $ {T^2 k} \gg 25\,\sqrt{|\delta_5 \cos 2\theta|}\,
\mathrm{MeV}^3$, eq. (\ref{case1I}) indicates that  $|\Delta_h(\la
E_{1,2},k)/\delta M^2|\gg 1$. In this high energy regime we find
that \bea \label{supsim} \delta \om_1^{(h)}(k,\la)-\delta
\om_2^{(h)}(k,\la) & \approx & \text{sign}\,(\delta
M^2)\Bigg[ A_e(k)-A_\mu(k)-B_e(k)+B_\mu(k)\Bigg] \\
&=& \text{sign}\,(\delta M^2) \;  \frac{28 \;  \pi^2}{45 \; \sqrt2}
\;
\frac{G_F \; k \;  T^4}{M_W^2} \\
& \simeq & 7.9 \times 10^{-15} \,\mathrm{eV} \; \text{sign}(\delta
M^2) \frac{k}{\mathrm{MeV}} \;
\left(\frac{T}{\mathrm{MeV}}\right)^4  \eea } \no therefore
neutrino oscillations are suppressed by a vanishingly small mixing
angle and the ratio of time scales (\ref{rattau}) becomes \be
\frac{\tau_v (k)}{\tau_m^{(h)} (k,\la)} \sim \Bigg|1+
\frac{10^{-3}}{|\delta_5|} \left(\frac{k
\,T^2}{\mathrm{MeV}^3}\right)^2 \Bigg|\,.\ee A considerable
\emph{speed-up} of oscillations occurs for $ k \; T^2 \gtrsim
100\,\mathrm{MeV}^3 $ since then $ \tau_m^{(h)} (k,\la) \ll \tau_v
(k) $. In this case, off-resonance flavor mixing is suppressed not
only by a small mixing angle in the medium but also by a rapid
decoherence and dephasing of the oscillations.

\item{ {\bf case II}: $\omega_{1,2}(k,\la) = \la \; (k +
\frac{M_{1,2}^2}{2k})~~;~~\la =+1, h=+1$ and $\la =-1, h=-1$: \\

The results of section \ref{resonance} (see eq. (\ref{caseII1}))
indicate  that   in this case  $|\Delta_h(\la E_{1,2},k)/\delta
M^2|\ll 1$, corresponding to the mixing angle in the medium being
the same as in the vacuum. As a result, \bea \label{vacsim} \delta
\om_1^{(h)}(k,\la)-\delta \om_2^{(h)}(k,\la)& \approx &
-\frac{\overline{M}^{\,2}}{4 \;  k^2}\cos
2 \; \theta \,\Bigg[A_e(k)-A_\mu(k) +B_e(k)-B_\mu(k)\Bigg] \\
&=& -\frac{\overline{M}^{\,2}}{4 \;  k^2} \; \cos 2\theta  \;
\frac{14 \;
\pi^2}{45  \; \sqrt2} \; \frac{G_F \; k \;  T^4}{M_W^2} \\
& \simeq & -6.1 \times 10^{-29} \,\mathrm{eV}  \;  \cos 2\theta \,
\left(\frac{k}{\mathrm{MeV}}\right)^{-1} \;
\left(\frac{T}{\mathrm{MeV}}\right)^4 \; .
 \eea }
\noindent and the ratio of time scales (\ref{rattau}) becomes \be
\frac{\tau_v (k )}{\tau_m^{(h)} (k,\la)} \simeq \Bigg|1 -
\frac{10^{-17}\cos 2\theta}{\delta_5} \left(\frac{T
 }{\mathrm{MeV} }\right)^4 \Bigg| \approx 1 \ee
Therefore, for positive helicity neutrinos and negative helicity
\emph{antineutrinos} medium oscillations are the same as vacuum
oscillations both in the mixing angle as well as in the oscillation
time scales. In this regime of temperature positive helicity
neutrinos and antineutrinos are sterile in the sense that these do
not interact with the medium.

\end{itemize}

\subsubsection{$\mathbf{m_e,m_\mu \ll T \ll M_W}$}

\begin{itemize}

\item{ {\bf case I}:  $\omega_{1,2}(k,\la) = \la (k +
\frac{M_{1,2}^2}{2k})~~;~~\la =+1, h=-1 $ and $ \la =-1, h=+1 $: \\

This case describes negative helicity neutrinos and positive
helicity antineutrinos. Eq. (\ref{caseI2}) indicates that in this
case, $|\Delta_h(\la E_{1,2},k)/\delta M^2|\gg 1$ corresponding to
vanishing mixing angle in the medium. Therefore, we obtain \bea
\delta \om_1^{(h)}(k,\la)-\delta \om_2^{(h)}(k,\la)& \approx &
\text{sign}\,(\delta
M^2)\Bigg[ A_e(k)-A_\mu(k)-B_e(k)+B_\mu(k)\Bigg] \\
&=& \text{sign}(\delta M^2) \,\, \frac{2 \; G_F \;  k \; T^2}{3 \;
\sqrt2} \; \left(\frac{m_\mu}{M_W}\right)^2 \\
& \simeq & 9.6 \times 10^{-6} \,\mathrm{eV} \;  \text{sign}(\delta
M^2) \, \left(\frac{k}{\mathrm{MeV}}\right) \;
\left(\frac{T}{\mathrm{GeV}}\right)^2 \; .
 \eea }
 \noindent and the ratio of time scales is given by
\be \frac{\tau_v (k )}{\tau_m^{(h)} (k,\la)} \simeq \Bigg|1 +
\frac{10^{12} }{|\delta_5|} \left(\frac{kT
 }{\mathrm{GeV}^2 }\right)^2 \Bigg|\gg 1 \ee
 Hence there is a considerable \emph{speed-up} in the oscillation time
 scale in the medium. Again, in this case oscillations are strongly
suppressed not only
 by a vanishingly small mixing angle but also by the rapid dephasing
 in the medium.

\item{ {\bf case II}: $\omega_{1,2}(k,\la) = \la \;  (k +
\frac{M_{1,2}^2}{2k})~~;~~\la =+1, h=+1$ and $\la =-1, h=-1$: \\

This case describes positive helicity neutrinos negative helicity
\emph{antineutrinos}.  Eq. (\ref{caseII2}) shows that in this case
$|\Delta_h(\la E_{1,2},k)/\delta M^2|\ll 1$, the mixing angle in the
medium is the same as in the vacuum. We find for this case \bea
\delta \om_1^{(h)}(k,\la)-\delta \om_2^{(h)}(k,\la)& \approx &
-\frac{\overline{M}^{\,2}}{4 k^2}\cos
2\theta \,\Bigg[A_e(k)-A_\mu(k) +B_e(k)-B_\mu(k)\Bigg] \\
&=& \frac{\overline{M}^{\,2}}{4 \;  k^2} \; \cos 2\theta \,\,
\frac{2 \; G_F
 \; k \; T^2}{3 \; \sqrt2}\left(\frac{m_\mu}{M_W}\right)^2 \\
& \simeq & 1.5 \times 10^{-19} \,\mathrm{eV}  \;  \cos 2\theta \,
\left(\frac{k}{\mathrm{MeV}}\right)^{-1} \;
\left(\frac{T}{\mathrm{GeV}}\right)^2 \; .
 \eea }
 \noindent where we have taken $\overline{M} \sim 1\,\mathrm{eV}$.
The ratio of time scales in this case is given by \be \frac{\tau_v
(k )}{\tau_m^{(h)} (k,\la)} \sim \Bigg|1 + \frac{3\times 10^{-8}\cos
2\theta }{ \delta_5 } \left(\frac{ T
 }{\mathrm{GeV}  }\right)^2 \Bigg| \approx 1  \ee
Again in this temperature regime we find that positive helicity
neutrinos and negative helicity \emph{antineutrinos } are
\emph{almost}  ''sterile'' in the sense that   neither the mixing
angle nor the oscillation time scales receive substantial loop
corrections. Thus the combined analysis of mixing angle and
propagation frequencies in the medium in the temperature regime
under consideration indicates that in medium corrections for
positive helicity neutrinos and negative helicity antineutrinos are
very small. These degrees of freedom are effectively sterile in that
their dynamics is (almost) the same as in the vacuum.
\end{itemize}

\section{Laplace Transform and Real-Time Evolution}\label{sec:laplace}

The main purpose to obtain the Dirac equation in real time is to
study the oscillations of neutrinos in the medium as an initial
value problem. As described in section \ref{sec:effective} this is
achieved by adiabatically switching the sources
$\eta,\overline{\eta}$ from $t = -\infty$ and switching them off at
$t=0$. The adiabatic switching on of the sources induces an
expectation value, which evolves in the absence of sources for
$t>0$, after the external source has been switched off. It is
convenient to write the effective Dirac eq. (\ref{eqnofmot}) in
terms of spatial Fourier transforms. Using the results of the
appendix (\ref{RetSEn}) we find \be \left[\left( i\gamma^0
\frac{\partial}{\partial t}-\vec{\gamma}\cdot \vec{k} \right)
\,\delta_{ab}-M_{ab}+\Sigma^{tad}_{ab}L\right]\,\psi_b (\vk,t) +
 \int_{-\infty}^t dt' \;  \Sigma_{ab}(\vk,t-t') \; L  \; \psi_b(\vk,t') =
- \eta_a(\vk,t), \label{eqnofmotft}\ee \noindent where the results
of appendix (\ref{RetSEn}) yield \be \Sigma (\vk,t-t') =
i\,\int_{-\infty}^\infty \frac{dk_0}{\pi} \;
\mathrm{Im}\Sigma(\vk,k_0) \;  e^{-ik_0(t-t')} \quad , \quad
\Sigma(\vk,k_0)=\Sigma_W(\vk,k_0)+\Sigma_Z(\vk,k_0)\label{selfa}\ee
For an external Grassmann valued source adiabatically switched on at
$t = -\infty$ and off at $t=0$ \be \eta_a(\vk,t) = \eta_a(\vk,0) \;
e^{\epsilon\,t} \; \theta(-t)\quad , \quad \epsilon \rightarrow 0^+
\; . \label{grasour} \ee It is straightforward to confirm that the
solution of the Dirac eq. (\ref{eqnofmotft}) for $t  < 0$ is given
by \be  \psi_a(\vk,t<0) =  \psi_a(\vk, 0) \; e^{\epsilon\,t} \; .
\label{solless} \ee Inserting this ansatze into the equation
(\ref{eqnofmotft}) it is straightforward to check that it is indeed
a solution with a linear relation between $ \psi_a(\vk, 0) $ and $
\eta_a(\vk,0) $. This relation  can be used to obtain $ \psi_a(\vk,
0) $ from $ \eta_a(\vk,0) $, or alternatively, for a given initial
value of the field at $t=0$ to find the source $ \eta_a(\vk,0) $
that prepares this initial value. For $ t>0 $ the source term
vanishes, the non-local integral in eq.(\ref{eqnofmotft}) can be
split into an integral from $ t=-\infty $ to $ t= 0$ plus an
integral from $ t=0 $ to $ t $. In the first integral corresponding
to $ t<0 $ we insert the solution eq.(\ref{solless}) and obtain the
following equation valid for $ t>0 $ \bea &&\left[\left( i\gamma^0
\frac{\partial}{\partial t}-\vec{\gamma}\cdot \vec{k} \right)
\,\delta_{ab}-M_{ab}+\Sigma^{tad}_{ab}L\right]\,\psi_b (\vk,t) +
 \int_{0}^t dt' \;  \Sigma_{ab}(\vk,t-t') \; L  \; \psi_b(\vk,t')
\nonumber \\&=&-\int_{-\infty}^{+\infty} \frac{dk_0}{\pi} \;
\frac{\mathrm{Im}\Sigma_{ab}(\vk,k_0)}{k_0}
 \; e^{-ik_0 t }\, L\,\psi_b(\vk, 0).\label{eqntgreat0}\eea

This equation can be solved by Laplace transform. Introduce the
Laplace transforms \be \label{laplatrafo} \widetilde{\psi}(\vk,s) =
\int_0^\infty dt  \; e^{-st}  \; \psi(\vk,t ) \quad , \quad
\widetilde{\Sigma}(\vk,s) = \int_0^\infty dt  \;  e^{-st} \;
\Sigma(\vk,t ) = \int_{-\infty}^{+\infty} \frac{dk_0}{\pi}\,
\frac{\mathrm{Im}\Sigma(\vk,k_0)}{k_0-is} \; , \ee \noindent where
we have used eq.(\ref{selfa}) to obtain the Laplace transform of the
self-energy, which leads to the analyticity relation (see Eq.
(\ref{retSEfin2})), \be \widetilde{\Sigma}(\vk,s) =
\Sigma(\vk,\omega=is-i\epsilon)  \; . \label{analit} \ee In terms of
Laplace transforms the equation of motion becomes the following
algebraic equation \be \left[ \left( i\gamma^0 s-\vec{\gamma}\cdot
\vec{k}  \right) \,\delta_{ab}-M_{ab}+\Sigma^{tad}_{ab}\,L   +
\widetilde{\Sigma}_{ab}(\vk,s)\,L \right]\widetilde{\psi}_b(\vk,s) =
i\left\{\gamma^0\;\delta_{ab}+\frac{1}{is}\left[
\widetilde{\Sigma}_{ab}(\vk,s)-\widetilde{\Sigma}_{ab}(\vk,0)\right]
\,L\right\}\psi_b(\vk,0)   \; . \label{lapladirac} \ee Consistently
with the expansion of the self-energy in frequency and momentum up
to order $\omega/M_W$, and using eq.(\ref{analit}), we replace the
expression in the bracket in eq.(\ref{lapladirac}) by \be
\frac{1}{is}\left[\widetilde{\Sigma}_{ab}(\vk,s)-
\widetilde{\Sigma}_{ab}(\vk,0)\right]\Big|_{s=0} =\frac{\partial
\Sigma (\vk,\omega)}{\partial \omega}\Big|_{\omega=0} \equiv
\Sigma'(\vk,0)   \; . \ee Using the representation (\ref{ReSig}) for
the real part of the self-energy, the  eq. (\ref{lapladirac}) can be
written as
 \be \left[ \left(  \gamma^0 \,i \; s-\vec{\gamma}\cdot
 \vk
\right) \,\mathds{1}-\mathds{M}+\gamma^0
\widetilde{\mathds{A}}(s)\,L -  \vec{\ga}\cdot\uvk \,\mathds{B}(
k)\,L \right]\widetilde{\psi}_b(\vk,s) =  i \; \gamma^0
\left[\mathds{1}+\mathds{A}'(0) \; L\right]\,\psi_b(\vk,0)   \; ,
\ee
 \noindent where
 \be \widetilde{\mathds{A}}(s)= \mathds{A}(\omega=is)~~;~~\mathds{A}'(0)=
 \frac{d\,\mathds{A}}{d\omega}\Big|_{\omega=0}  \; .\label{laplaA}
\ee The real time evolution is obtained by the inverse Laplace
transform, \be \psi(\vk,t) = \int_{\Gamma} \frac{ds}{2\pi i}\;
\widetilde{\psi}(\vk,s)\; e^{st}   \; ,\label{inverlapla} \ee
\noindent where $\Gamma$ is the Bromwich contour in the complex
$s$ plane running parallel to the imaginary axis to the right of
all the singularities of the function $\widetilde{\psi}(\vk,s)$
and closing on a large semicircle to the left. We now follow the
same steps as in section \ref{sec:diagonal}, namely projecting
onto right and left components and onto helicity eigenstates.
After straightforward manipulations we arrive at the following set
of equations \be \left[-(s^2+k^2)  \; \mathds{1}+(is-\hat{h}(\uvk)
k)(\widetilde{\mathds{A}}+\hat{h}(\uvk) \mathds{B})-\mathds{M}^2
\right]\widetilde{\psi} _L(\vk,s)= i \; \gamma^0  \;  \mathds{M}
\; \psi _R(\vk,0)+i(is-\hat{h}(\uvk) k) \; \mathds{D}\;\psi
_L(\vk,0) \; , \label{lapL} \ee \noindent where $\mathds{D}=
\mathds{1}+\mathds{A}'(0) $, and \be\widetilde{\psi}_R(\vk,s) =
-\frac{is+\hat{h}(\uvk)k}{s^2+k^2}\left[\mathds{M}  \; \gamma^0 \;
\widetilde{\psi}_L(\vk,s)+i   \; \psi_R(\vk,0)\right] \; . \ee We
now follow the same steps as above to separate the Dirac and
flavor structures by introducing the flavor doublets
$\widetilde{\varphi}( \vk,s), \; \widetilde{\xi}( \vec{k},s)$
which are the Laplace transform of the flavor doublets $
\varphi(\vk,t), \; \xi(\vk,t) $ introduced in the expansion of the
Dirac spinors in eqs.(\ref{geneL})-(\ref{geneR}), projecting onto
the Weyl spinors eigenstates of helicity, the above equations
become \be \left[-(s^2+k^2)\mathds{1}+(is-h
k)(\widetilde{\mathds{A}}+h \mathds{B})-\mathds{M}^2
\right]\widetilde{\varphi}^{(h)} (\vk,s)= -i   \; \mathds{M}
\,\xi^{(h)}(\vk,0)+i  \; (is-h k)  \; \mathds{D}\;\varphi^{(h)}
(\vk,0) \label{lapLproy}   \; ,\ee \be\widetilde{\xi}^{(h)}
(\vk,s) = -\frac{is+hk}{s^2+k^2}\left[-\mathds{M}\,
\widetilde{{\varphi}}^{(h)} (\vk,s)+i   \;  \xi^{(h)}
(\vk,0)\right]\label{xitil}   \; . \ee

The solution to eq. (\ref{lapLproy}) is obtained as \be
\widetilde{\varphi}^{(h)} (\vk,s)=
\widetilde{\mathbb{S}}^{(h)}(k,s)\,\Bigg[-i \mathds{M} \;
\xi^{(h)}(\vk,0)+i(i \; s-h \;  k) \mathds{D}\,\varphi^{(h)} (\vk,0)
\Bigg], \label{laplasol} \ee \no where the propagator is given by
\be \label{propa} \widetilde{\mathbb{S}}^{ h }(k,s) =
\frac{1}{\alpha^2_{h}(k,s)-\beta^2_{h}(k,s)} \Bigg(
          \begin{array}{cc}
           \alpha_{h}(k,s)+\beta(k,s)\cos 2\theta^{(h)}_m &
-\beta_{h}(k,s)\sin 2\theta^{(h)}_m  \\
             -\beta_{h}(k,s)\sin 2\theta^{(h)}_m  & \alpha_{h}(k,s)-
\beta_{h}(k,s)\cos 2\theta^{(h)}_m  \\
          \end{array}
        \Bigg), \ee
\no in which it will prove convenient to introduce the following
quantities \be\label{alfa} \alpha_{ h }(k,s) = \Bigg[ \omega^2-k^2
+\frac12 \; S_h(\omega,k)-\frac12 \;
(M^2_1+M^2_2)\Bigg]_{\omega=is-i\epsilon} \; , \ee \be \label{beta}
\beta_{ h }(k,s) =  \left. \frac12 \; \delta M^2 \left[\left(\cos
2\theta-\frac{\Delta_h (\omega,k)}{\delta
M^2}\right)^2+\sin^22\theta
\right]^{\frac12}\right|_{\omega=is-i\epsilon}\;. \ee

The inverse Laplace transform eq.(\ref{inverlapla}) can be done
straightforwardly, the singularities of $
\tilde{\varphi}^{(h)}(\vk,s) $ in the complex $s$ plane are
determined by the singularities of the propagator $
\widetilde{\mathbb{S}}^{(h)}(k,s) $. Up to the order in weak
interactions considered here, these singularities are isolated poles
along the imaginary axis at the positions $ s = -i \;
\omega^{(h)}_a(k,\pm) $ given by eq. (\ref{omegaa})-(\ref{delom2}).
As a relevant example of the real time evolution, let us consider
that the initial state corresponds to a wave-packet of left-handed
electron neutrinos of arbitrary helicity $h$, with no muon
neutrinos. This could, for example, be the case relevant for
nucleosynthesis in which a neutron beta decays at the initial time.
In this case \be \varphi^{(h)} (\vk,0) = \nu^{(h)}_e(\vk) \left(
          \begin{array}{c}
            1 \\
            0 \\
          \end{array}\right) \quad , \quad
\xi^{(h)}(\vk,0)= \left(
          \begin{array}{c}
            0 \\
            0 \\
          \end{array}\right) \; , \label{inistate}\ee
 and we find
 \be
\widetilde{\varphi}^{(h)} (\vk,s) = \nu^{(h)}_e(\vk)\;
 \frac{i(i\; s-h \; k)[1+A'_e(0)]}{\alpha^2_{h}(k,s)-\beta^2_{h}(k,s) }\,\left(
          \begin{array}{c}
            \alpha_{h}(k,s)+\beta_{h}(k,s) \,\cos(2\theta^{(h)}_m)\\
            -\beta_{h}(k,s) \,\sin(2\theta^{(h)}_m ) \\
          \end{array}\right) \; , \ee
 \be
\widetilde{\xi}^{(h)} (\vk,s) = -\nu^{(h)}_e(\vk)\,
 \frac{i [1+A'_e(0)]}{\alpha^2_{h}(k,s)-\beta^2_{h}(k,s) } \; \mathds{M}\;
\left( \begin{array}{c}
            \alpha_{h}(k,s)+\beta_{h}(k,s) \,\cos(2\theta^{(h)}_m)\\
            -\beta_{h}(k,s) \,\sin(2\theta^{(h)}_m ) \\
          \end{array}\right)\; .
\ee In order to avoid cluttering the notation, we have not included
the frequency argument in the mixing angle in the medium
$\theta^{(h)}_m$ but such dependence should be understood
throughout.

The term $ [\alpha_h(k,s)-\beta_h(k,s)]^{-1} $ features poles at $
s= -i\omega_1(k,\pm) $ and the term $
[\alpha_h(k,s)+\beta_h(k,s)]^{-1} $ features poles at $ s=
-i\omega_2(k,\pm) $.

We neglect terms of order $ G_F\;  T^4/M^2_W \sim (T/M_W)^4 \ll 1 $
since in the regime in which the approximations are valid $ T\ll M_W
$. The residues of these poles are respectively $
2\,\omega_{1,2}(k,\pm) $
 and the  inverse Laplace transform  yield within these approximations,
\be  \label{phioft} \varphi^{(h)}(\vk,t) =
\nu^{(h)}_e(\vk)\sum_{\lambda=\pm} \Bigg[
\frac{\omega^{(h)}_1(k,\lambda)-h\,k}{4\,\omega^{(h)}_1(k,\lambda)
}\, \left(
          \begin{array}{c}
            1+C^{(h)}_{1,\lambda}\\
             -S^{(h)}_{1,\lambda}  \\
          \end{array}\right)  \,
e^{-i\omega^{(h)}_1(k,\lambda)\,t}+
\frac{\omega^{(h)}_2(k,\lambda)-h\,k}{4\,\omega^{(h)}_2(k,\lambda)}\;
\left(           \begin{array}{c}
            1-C^{(h)}_{2,\lambda}\\
             S^{(h)}_{2,\lambda} \\
          \end{array}\right) \, e^{-i\omega^{(h)}_2(k,\lambda)\,t} \Bigg] \; ,
          \ee
\be \label{xioft} \xi^{(h)}(\vk,t) = -
\nu^{(h)}_e(\vk)\sum_{\lambda=\pm} \Bigg[
\frac{\mathds{M}}{4\,\omega^{(h)}_1(k,\lambda) }\, \left(
          \begin{array}{c}
            1+C^{(h)}_{1,\lambda}\\
             -S^{(h)}_{1,\lambda}  \\
          \end{array}\right)  \,
e^{-i\omega^{(h)}_1(k,\lambda)\,t}+
\frac{\mathds{M}}{4\,\omega^{(h)}_2(k,\lambda)}\,\left(
          \begin{array}{c}
            1-C^{(h)}_{2,\lambda}\\
             S^{(h)}_{2,\lambda} \\
          \end{array}\right) \, e^{-i\omega^{(h)}_2(k,\lambda)\,t} \Bigg] \; .
          \ee
For economy of notation, we   introduce  the following shorthand \be
C^{(h)}_{a,\lambda} \equiv
\cos\left[2\theta^{(h)}_m(\omega^{(h)}_a(k,\lambda)) \right] \quad ,
\quad S^{(h)}_{a,\lambda} \equiv
\sin\left[2\theta^{(h)}_m(\omega^{(h)}_a(k,\lambda))
\right]\label{S1lam}\ee \noindent for $a=1,2$, helicity components
$h=\pm$ and positive and energy components $\lambda =\pm$.

The above expressions yield a direct comparison with the usual
oscillation formulae in the literature. To leading order we set $
\omega_{1,2}(k,\lambda) \approx \lambda k \; (1+M^2_{1,2}/2k^2) $ in
the prefactors in the expressions above thereby neglecting terms of
order $G_F$.  Since the dependence of the mixing angle on the
frequency and momentum appears at order $ G_F $, we can set $
\omega_{1,2}(k,\lambda) \approx \lambda \;  k $ in the argument of
the mixing angles, therefore to leading order $
C^{(h)}_{1,\lambda}=C^{(h)}_{2,\lambda}\equiv C^{(h)}_{ \lambda} $
and $ S^{(h)}_{1,\lambda}=S^{(h)}_{2,\lambda}\equiv S^{(h)}_{
\lambda}
 $. With these approximations, for an initial left handed electron
state of helicity $ h=\mp $ we find \bea \label{varm}\varphi^{ -
}(\vk,t) &=& \nu^{-}_e(\vk)\Bigg[ \frac12 \left(
          \begin{array}{c}
            1+C^{ - }_{ +}\\
             -S^{ - }_{ +}  \\
          \end{array}\right)  \, e^{-i\omega^{ - }_1(k,+)\,t} + \frac12 \left(
          \begin{array}{c}
            1-C^{ - }_{ +}\\
             S^{ - }_{ +}  \\
\end{array}\right)  \, e^{-i\omega^{ - }_2(k,+)\,t} \nonumber          \\
&+& \frac{M^2_1}{8 k^2} \left(
          \begin{array}{c}
            1+C^{ - }_{ -}\\
             -S^{ - }_{ -}  \\
\end{array}\right)  \, e^{- i\omega^{ - }_1(k,-)\,t} +\frac{M^2_2}{8 k^2} \left(
          \begin{array}{c}
            1-C^{ - }_{ -}\\
             S^{ - }_{ -}  \\
          \end{array}\right)  \, e^{ -i\omega^{ - }_2(k,-)\,t}
          \Bigg] \; ,
          \eea
\bea \label{varp} \varphi^{ + }(\vk,t) &=& \nu^{+}_e(\vk)\Bigg[
\frac{M^2_1}{8 k^2} \left(
          \begin{array}{c}
            1+C^{ + }_{ +}\\
             -S^{ + }_{ +}  \\
\end{array}\right)  \, e^{-i\omega^{ + }_1(k,+)\,t} + \frac{M^2_2}{8 k^2} \left(
          \begin{array}{c}
            1-C^{ + }_{ +}\\
             S^{ + }_{ +}  \\
\end{array}\right)  \, e^{-i\omega^{+ }_2(k,+)\,t} \nonumber          \\
&+&\frac{ 1}{2 } \left(
          \begin{array}{c}
            1+C^{ + }_{ -}\\
             -S^{ + }_{ -}  \\
\end{array}\right)  \, e^{ -i\omega^{+ }_1(k,-)\,t} +\frac{1}{2 } \left(
          \begin{array}{c}
            1-C^{ +}_{ -}\\
             S^{ +}_{ -}  \\
          \end{array}\right)  \, e^{- i\omega^{ + }_2(k,-)\,t}
          \Bigg] \; .
          \eea
The exponentials $ e^{\mp i \omega(k)\,t} $ correspond to positive
energy $ (-) $  neutrino and positive energy $ (+) $  antineutrino
components respectively. Therefore, the expressions above reveal
that for negative helicity the relevant components in the
relativistic limit correspond to positive energy neutrinos, while
for positive helicity they correspond to positive energy
\emph{antineutrinos}.

The upper component of the expressions above correspond to
wavepackets  of negative  and positive helicity respectively for a
left handed electron neutrino, namely $ \nu^{\mp}_e(\vk,t) $ while
the lower components correspond  to a left handed \emph{muon}
neutrino, namely $ \nu^{\mp}_\mu(\vk,t) $.

For the right handed components $\xi$ the leading order can be
obtained by setting $\mathds{M}\approx \overline{M}\,\mathds{1}$
thereby neglecting terms of order $\delta M^2/\overline{M}^{\,2}$.
This approximation yields, \bea \label{xis} \xi^{ \mp }(\vk,t)  &=&
- \nu^{\mp}_e(\vk) \; \frac{\overline{M}}{4 \; k} \; \Bigg[ \left(
          \begin{array}{c}
            1+C^{ \mp }_{ +}\\
             -S^{ \mp }_{ +}  \\
          \end{array}\right)  \, e^{-i\omega^{ \mp }_1(k,+)\,t} +   \left(
          \begin{array}{c}
            1-C^{ \mp }_{ +}\\
             S^{ \mp }_{ +}  \\
\end{array}\right)  \, e^{-i\omega^{ \mp }_2(k,+)\,t} \nonumber          \\
&& -  \left(
          \begin{array}{c}
            1+C^{ \mp }_{ -}\\
             -S^{ \mp }_{ -}  \\
          \end{array}\right)  \, e^{- i\omega^{ \mp }_1(k,-)\,t} -  \left(
          \begin{array}{c}
            1-C^{ \mp }_{ -}\\
             S^{ \mp }_{ -}  \\
          \end{array}\right)  \, e^{- i\omega^{ \mp }_2(k,-)\,t}
          \Bigg] \; .
          \eea
The upper and lower components correspond to wavepackets for right
handed negative and positive helicity electron and muon neutrinos
respectively.

From the expression  (\ref{varm})  we can read off the probability
for relativistic left handed, negative helicity electron and muon
neutrinos as a function of time, \bea |\nu^{-}_{e,L}(\vk,t)|^2  & =
& |\nu^{-}_e(\vk)|^2\left\{1-\sin^2[2 \; \theta^{-}_m(k)] \;
\sin^2\left(\frac12  \; \Delta \omega^-(k,+) \; t \right)+
\mathcal{O}\left(\frac{M^2_{1,2}}{k^2}\right)\right\}  \; , \label{probel}\\
|\nu^{-}_{\mu,L}(\vk,t)|^2  & =  & |\nu^{-}_e(\vk)|^2 \left\{
\sin^2[2 \; \theta^{-}_m(k)] \; \sin^2\left( \frac12 \; \Delta
\omega^-(k,+)\; t
\right)+\mathcal{O}\left(\frac{M^2_{1,2}}{k^2}\right)\right\}
\label{probmu} \; . \eea The probability for left handed
relativistic \emph{positive} helicity electron and muon
antineutrinos as a function of time  are read off from
eq.(\ref{varp}) \bea |\nu^{+}_{e,L}(\vk,t)|^2  & =  &
|\nu^{+}_e(\vk)|^2\left\{ 1-\sin^2[2 \; \theta^{+}_m(-k)] \;
\sin^2\left(\frac12  \; \Delta \omega^+(k,-) \; t \right)+
\mathcal{O}\left(\frac{M^2_{1,2}}{k^2}\right)\right\} \label{probpel} \; , \\
|\nu^{+}_{\mu,L}(\vk,t)|^2  & =  & |\nu^{+}_e(\vk)|^2 \left\{
\sin^2[2 \; \theta^{+}_m(-k)] \;  \sin^2\left( \frac12 \; \Delta
\omega^+(k,-) \; t \right)+
\mathcal{O}\left(\frac{M^2_{1,2}}{k^2}\right)\right\}
\label{probpmu} \; , \eea \noindent where \be \Delta
\omega^{\pm}(k,\pm) = \frac{\delta M^2}{2k}+\delta
\omega^{\pm}_1(k,\pm)-\delta \omega^{\pm}_2(k,\pm) \; .
\label{Delomp} \ee The corrections $\delta \omega^{\pm}_a(k,\pm)$
had been studied in detail in section \ref{osfreq} above. Finally,
eq. (\ref{xis}) determines the probabilities of finding \emph{right
handed} positive and negative helicities neutrinos as a function of
time. This equation makes manifest that this probability is
suppressed by a factor $ \overline{M}/k $ with respect to the left
handed component, indeed it is the mass term that is responsible for
generating a right handed component from a left handed one and must
therefore be suppressed by one power of the ratio $ M/k $. For a
typical neutrino momentum $ k \gtrsim 1\,\mathrm{MeV} $ this
suppression factor is of order $ 10^{-6}$.
Eqs.(\ref{phioft})-(\ref{xioft}) provide a complete field
theoretical description of oscillations in \emph{real time}.
Eqs.(\ref{probel})-(\ref{probmu}) are obviously reminiscent of the
familiar oscillation equations obtained in the simplified quantum
mechanical two level system, however there are some important
aspects that must be highlighted, namely, the field theoretical
formulation introduced here led directly to these oscillation
formulae in terms of the mixing angles in the medium and the correct
oscillation frequencies that include the quantum loop corrections.
Furthermore, the oscillation formulae obtained above reveal the
nature of the approximations and allow a consistent inclusion of
higher order effects as well as describe the oscillation of
\emph{all} helicity components as well as the dynamics of the right
handed component. The usual oscillation formula obtained within the
single particle quantum mechanical description emerge cleanly in
suitable limits and the nature of the corrections is readily
manifest.

\section{Conclusions,  and further
questions:}\label{conclu}

We have provided a systematic treatment of neutrino oscillations and
mixing directly from quantum field theory in real time in a regime
of temperature and density relevant for early Universe cosmology
prior to big bang nucleosynthesis. While we have focused on two
flavors (electron and muon) of Dirac neutrinos the formulation can
be generalized straightforwardly to more flavors and to
Majorana-Dirac mass matrices.

We have obtained the medium corrections to the dispersion relations
and mixing angles of propagating neutrinos. Implementing methods
from real time non-equilibrium quantum field theory at finite
temperature and density we have systematically obtained the
equations of motion for the neutrino field and studied the real time
evolution as an initial value problem. The major advantage of this
approach, as compared to the usual approach based on single particle
quantum mechanics  is that it consistently and systematically
includes the medium corrections to the dispersion relations and
mixing angles \emph{directly} into the real time evolution and
treats left and right handed fields and both helicity components on
equal footing. We have argued that collisional relaxation yields
thermalization of neutrinos in \emph{flavor eigenstates} for
temperature $T \gtrsim 5-10\,\mathrm{Mev}$ for which the relaxation
time scale via weak interactions is shorter than the oscillation
time scale. Assuming the validity of this argument, we obtained the
neutrino self-energies up to one-loop including the asymmetries from
leptons, neutrinos, hadrons and quarks, as well as  non-local (in
space-time) terms arising from the expansion of the self-energy loop
in the external frequency and momentum. We have consider these
non-local terms up to leading order in $\omega/M_W~;~k/M_W$ since
these terms are of the same order of or larger than the contribution
from the asymmetries for $T \gtrsim 5 \, \mathrm{MeV}$ \emph{if} all
asymmetries are of the same order as the baryon asymmetry. This is
yet another indication that this is an important temperature regime
in the early Universe.

Our main results are summarized as follows:

\begin{itemize}
\item{Implementing the methods from non-equilibrium real-time
field theory at finite temperature and density we obtained the
equations of motion for the neutrino fields in linear response. This
formulation includes consistently the self-energy loop corrections
to the dispersion relations and mixing angles in the medium and
treat left and right handed fields with both helicity components on
equal footing.  }

\item{ We have focused on a temperature regime prior to
nucleosynthesis $T \gtrsim 5-10\,\mathrm{MeV}$ in which we argued
that neutrinos are thermalized as flavor eigenstates. We studied two
different temperature regimes: $m_e \ll T \ll m_\mu$ within which we
have shown that there is the possibility of resonant oscillations of
test neutrinos, and $m_e, m_\mu \ll T \ll M_W$ within which the
mixing angle for active neutrinos effectively vanishes.  }

\item{An expansion of the self-energy in terms of the neutrino frequency and momentum is carried
out to lowest order in $\omega/M_W~;~k/M_W$ thus extracting the
leading \emph{non-local} (in space-time) contributions. We find a
\emph{new} contribution which cannot be identified with an effective
potential. The mixing angles and propagation frequencies in the
medium are found to be \emph{helicity dependent}. It has been
recently pointed out\cite{bento} that a neutrino helicity asymmetry
could be a very important ingredient in successful see-saw models of
thermal leptogenesis.}

\item{ If the lepton and quark
asymmetries are of the same order as the baryon asymmetry in the
early Universe, we have shown that the non-local  (in space-time)
terms in the self-energies dominate over the asymmetry for typical
energies of neutrinos in the plasma for $T \gtrsim 3-5\,
\mathrm{MeV}$.  }

\item{The oscillation time scale in the medium is
\emph{slowed-down } near the resonance, becoming substantially
\emph{longer} than in the vacuum for small vacuum mixing angle. For
high energy neutrinos off-resonance the mixing angle becomes
vanishingly small and the oscillation time scale \emph{speeds-up} as
compared to the vacuum. At high temperature, in the region $T\gg
m_e,m_\mu$ the mixing angle for active neutrinos effectively
vanishes and there is a considerable \emph{speed-up} of
oscillations, which are then suppressed by a vanishingly small
mixing angle and a rapid dephasing. }

\item{ We have obtained the general equations of motion for
initially prepared wave packets of neutrinos of arbitrary chirality
and helicity. These equations reduce to the familiar oscillation
formulae for ultrarelativistic negative helicity neutrinos, but with
the bonus that they consistently include the mixing angles and the
oscillation frequencies in the medium. These equations not only
yield the familiar ones but also quantify the magnitude of the
corrections. Furthermore these equations also describe the evolution
of right handed neutrinos (of either helicity) which is a
consequence of a non-trivial mass matrix and usually ignored in the
usual formulation. }

\end{itemize}

{\bf Further questions:} The next step in the program is to provide
a solid assessment of the reliability of the argument that suggests
that for $T \gtrsim 10 \,\mathrm{MeV}$   neutrinos thermalize as
flavor eigenstates. The phenomenon of \emph{slow-down} of
oscillations near a resonance found in this work, suggests that this
argument is consistent at least near resonances.
 We are currently studying  the relaxational dynamics by including two
loop corrections to the self-energy which account for collisional
processes in the medium. We are also studying how to obtain the
kinetic equations that describe simultaneously oscillations and
relaxation systematically within a field theory approach that
accounts for the subtle aspects of the Fock representation necessary
to understand the concept of the flavor distribution
functions\cite{DanMan}.

\begin{acknowledgments} D.B. thanks the US NSF for the support under
grant PHY-0242134 and A. Dolgov for discussions. C.M.Ho.
acknowledges the support through the Andrew Mellon Foundation.
\end{acknowledgments}

\appendix

\section{Real-time propagators and self-energies.}

\subsection{Fermions}

Consider a generic fermion field $f(\vx,t)$ of mass $m_f$. The
Wightmann and Green's functions at finite temperature are given as
\bea i \;  S^>_{\alpha,\beta} (\vx-\vx',t-t') & = & \langle
f_{\alpha}(\vx,t) \overline{f}_{\beta}(\vx',t')\rangle =
\frac{1}{V}\sum_{\vp}e^{i \vp\cdot(\vx-\vx')}
\;iS^>_{\alpha,\beta}(\vp,t-t'),
\label{sgreat}\\
i \;  S^<_{\alpha,\beta} (\vx-\vx',t-t') & = & -\langle
\overline{f}_{\beta}(\vx',t') f_{\alpha}(\vx,t) \rangle =
\frac{1}{V}\sum_{\vp}e^{i \vp\cdot(\vx-\vx')} \;i \;
S^<_{\alpha,\beta}(\vp,t-t'), \label{sless}\eea \noindent where $
\alpha,\beta $ are Dirac indices and $ V $ is the quantization
volume.

The real-time Green's functions along the forward $ (+) $ and
backward $ (-) $ time branches are given in terms of these Wightmann
functions as \bea \langle f^{(+)}_{\alpha}(\vx,t)
\overline{f}^{(+)}_{\beta}(\vx',t')\rangle & = & i \;
S^{++}(\vx-\vx',t-t')= i \;  S^>(\vx-\vx',t-t') \; \Theta(t-t')+ i
\; S^<(\vx-\vx',t-t') \; \Theta(t'-t) \; ,
\label{Spp} \\
\langle f^{(+)}_{\alpha}(\vx,t)
\overline{f}^{(-)}_{\beta}(\vx',t')\rangle & = & i \;
S^{+-}(\vx-\vx',t-t')= i \;  S^<(\vx-\vx',t-t') \; .\eea At finite
temperature $T$, it is straightforward to obtain these correlation
functions by expanding the free fermion fields in terms of Fock
creation and annihilation operators and massive spinors. In a $CP$
asymmetric medium, the chemical potential $\mu_{f}$ for the fermion
$f$ is non-zero. Particles and anti-particles obey the following
Fermi-Dirac distribution functions respectively \be
N_{f}(p_0)=\frac{1}{e^{(p_0-\mu_{f})/T}+1} \quad , \quad
\bar{N}_{f}(p_0)=\frac{1}{e^{(p_0+\mu_{f})/T}+1} \; . \ee The
fermionic propagators are conveniently written in a dispersive form
\bea i \; S^>_{\alpha,\beta}(\vp,t-t') &=& \int_{-\infty}^{\infty}
dp_0 \; \rho^>_{\alpha,\beta}(\vp,p_0) \;  e^{-ip_0(t-t')} \; ,
\label{Sgreatdis}\\
i \; S^<_{\alpha,\beta}(\vp,t-t') & = & \int_{-\infty}^{\infty} dp_0
\; \rho^<_{\alpha,\beta}(\vp,p_0) \;  e^{-ip_0(t-t')} \; ,
\label{Slessdis}\eea \noindent where we have \bea
\rho^>_{\alpha,\beta}(\vp,p_0)&=& \frac{\gamma^0\,p_0 -
\vec{\gamma}\cdot \vp+m_f}{2p_0}  \; [1-N_f(p_0)] \; \delta(p_0-\p)
+ \frac{\gamma^0\,p_0 - \vec{\gamma}\cdot \vp+m_f}{2p_0} \;
\bar{N}_f(-p_0) \; \delta(p_0+\p) \; , \\
\rho^<_{\alpha,\beta}(\vp,p_0)&=& \frac{\gamma^0\,p_0 -
\vec{\gamma}\cdot \vp+m_f}{2p_0} \;  N_f(p_0) \; \delta(p_0-\p) +
\frac{\gamma^0\,p_0 - \vec{\gamma}\cdot \vp+m_f}{2p_0} \;
[1-\bar{N}_f(-p_0)] \; \delta(p_0+\p) \; , \eea \noindent with $
\p=\sqrt{|\vec{p}\,|^2+m_f^2} $. Using the relation $
\bar{N}_f(-p_0)=1-N_f(p_0) $, we can write \bea
\rho^>_{\alpha,\beta}(\vp,p_0) &=& [1-N_f(p_0)] \;
\rho^f_{\alpha,\beta}(\vp,p_0) \; , \label{rhogreat}\\
\rho^<_{\alpha,\beta}(\vp,p_0) & = & N_f(p_0) \;
\rho^f_{\alpha,\beta}(\vp,p_0) \; , \label{rholess} \eea \noindent
where the free fermionic spectral density $ \rho^f(\vp,p_0) $ is
given by \bea \rho^f(\vp,p_0)&=& \frac{\not\!{p}_+}{2\p} \;
\delta(p_0-\p)+
\frac{\not\!{p}_-}{2\p} \; \delta(p_0+\p) \; ,\label{rhofree}\\
{\not\!{p}_{\pm}} &=& \gamma^0\,\p \mp \vec{\gamma}\cdot \vp\,\pm
m_f \label{ppm}  \; . \eea
\subsection{Vector Bosons}

Consider a generic real vector boson field $ A_{\mu}(\vx,t) $ of
mass $M$. In unitary gauge, it can be expanded in terms of Fock
creation and annihilation operators of \emph{physical} states with
three polarizations as \be A^{\mu}(\vx,t)=
\frac{1}{\sqrt{V}}\sum_{\lambda} \sum_{\vk}
\frac{\epsilon^{\mu}_{\lambda}(\vk)}{\sqrt{2 \; W_k}}
\left[a_{\vk,\lambda} \; e^{-iW_k\, t} \; e^{i
\vk\cdot\vx}+a^{\dagger}_{\vk,\lambda} \; e^{iW_k\, t} \; e^{-i
\vk\cdot\vx}\right] ~~;~~ k^{\mu} \; \epsilon_{\mu,\lambda}(\vk) =0
\; , \ee \noindent where $W_k=\sqrt{|\vec{k}\,|^2+M^2}$ and
$k^{\mu}$ is \emph{on-shell} $k^{\mu}=(W_k,\vk)$. The three
polarization vectors are such that \be \sum_{\lambda=1}^3
\epsilon^{\mu}_{\lambda}(\vk) \; \epsilon^{\nu}_{\lambda}(\vk)=
P^{\mu \nu}(\vk)= - \left( g^{\mu \nu}- \frac{k^\mu
k^\nu}{M^2}\right) \; . \label{projector} \ee It is now
straightforward to compute the Wightmann functions of the vector
bosons in a state in which the physical degrees of freedom are in
thermal equilibrium at temperature $ T $. These are given by \bea
\langle A_{\mu}(\vx,t) A_{\nu}(\vx',t') \rangle & = &
i \; G^>_{\mu,\nu}(\vx-\vx',t-t'), \label{AAgreat} \\
\langle A_{\nu}(\vx',t') A_{\mu}(\vx,t)\rangle & = & i \;
G^<_{\mu,\nu}(\vx-\vx',t-t'), \label{AAless}\eea \noindent where
$G^{<,>}$ can be conveniently written as spectral integrals in the
form \bea i \; G^>_{\mu,\nu}(\vx-\vx',t-t') & = &
\frac{1}{{V}}\sum_{\vk}
e^{i\vk\cdot(\vx-\vx')}\int_{-\infty}^{\infty}dk_0 \;
e^{-ik_0(t-t')} \; [1+N_b(k_0)] \; \rho_{\mu \nu}(k_0,\vk) \; ,
\label{Ggreat}\\
i \; G^<_{\mu,\nu}(\vx-\vx',t-t') & = & \frac{1}{{V}}\sum_{\vk}
e^{i\vk\cdot(\vx-\vx')} \int_{-\infty}^{\infty}dk_0 \;
e^{-ik_0(t-t')} \; N_b(k_0) \; \rho_{\mu \nu}(k_0,\vk) \; ,
\label{Gless} \eea \noindent where \be N_b(k_0)=
\frac{1}{e^{k_0/T}-1} \; ,\label{BEdist} \ee \noindent and the
spectral density is given by \be \rho_{\mu \nu}(k_0,\vk) =
\frac{1}{2W_k}\left[P_{\mu \nu}(\vk) \; \delta(k_0-W_k)- P_{\mu
\nu}(-\vk) \; \delta(k_0+W_k)\right] \; .\label{vbspecdens} \ee In
terms of these Wightmann functions, the real-time correlation
functions along the forward and backward time branches are given by
\bea \langle A^{(+)}_{\mu}(\vx,t) A^{(+)}_{\nu}(\vx',t') \rangle & =
& iG^>_{\mu,\nu}(\vx-\vx',t-t') \; \Theta(t-t')+
i \; G^<_{\mu,\nu}(\vx-\vx',t-t') \; \Theta(t'-t), \label{App}\\
\langle A^{(+)}_{\mu}(\vx,t) A^{(-)}_{\nu}(\vx',t') \rangle & = & i
\; G^<_{\mu,\nu}(\vx-\vx',t-t') \; . \label{Apm} \eea For the
charged vector bosons, the correlation functions can be found simply
from those of the real vector boson fields described above by
writing the charged fields as linear combinations of \emph{two} real
fields $ A^{1,2} $, namely \be W^{\pm}_{\mu}(\vx,t) =
\frac{1}{\sqrt2} [A^{1}_{\mu}(\vx,t)\pm i A^2_{\mu}(\vx,t)] \; . \ee
It is straightforward to find the correlation function \be \langle
W^{+}_{\mu}(\vx,t)W^{-}_{\mu}(\vx',t') \rangle =
G^>_{\mu\nu}(\vx-\vx',t-t') \; , \ee \noindent and similarly for the
other necessary Wightmann and Green's functions.

\subsection{Retarded Self-Energies for Charged and Neutral Current
Interactions}\label{RetSEn}

\begin{figure}[h!]
\begin{center}
\includegraphics[height=3in,width=4in,keepaspectratio=true]{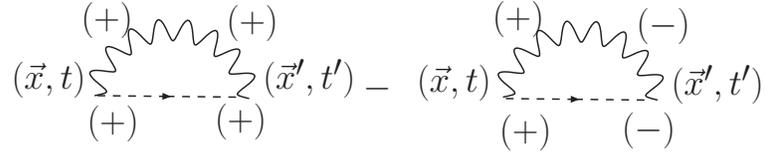}
\caption{Retarded self-energy for charged current interactions. The
wiggly line is a charged vector boson and the dashed line a lepton.
The labels $ (\pm) $ correspond to the forward ($ + $) and backward
($ - $) time branches. The corresponding propagators are $ i \;
S^{\pm,\pm}(\vx-\vx',t-t') $ and $ i \; G^{\pm
\pm}_{\mu\nu}(\vx-\vx',t-t') $ for leptons and charged bosons
respectively. } \label{retSE}
\end{center}
\end{figure}

The diagrams for the one-loop retarded self-energy from charged
current interactions are displayed in Fig. (\ref{retSE}). A
straightforward calculation yields for the charged current
contribution the following result \be
\Sigma^{CC}_{ret}(\vx-\vx',t-t') = \frac{ig^2}{2} \;  R \;
\gamma^{\mu}\left[i \; S^{++}(\vx-\vx',t-t') \; i \;
G^{++}_{\mu\nu}(\vx-\vx',t-t')-i \; S^<(\vx-\vx',t-t') \; i \;
G^{<}_{\mu\nu}(\vx-\vx',t-t')\right] \; \gamma^\nu \;  L \; ,
\label{retaSE} \ee \noindent with
$$
R=\frac{1+\gamma^5}{2} \quad , \quad L=\frac{1-\gamma^5}{2} \; .
$$
\no A similar result is obtained for the neutral current
contribution to the self-energy by simply replacing $ g/\sqrt2
\rightarrow g/2\cos\theta_w $ and $ M_W\rightarrow
M_Z=M_W/\cos\theta_w $.

Using the representation of the fermion and vector boson propagators
given above the retarded self-energy (\ref{retaSE}) can be written
as \be \Sigma_{ret}(\vx-\vx',t-t') = \frac{i}{V} \; \sum_{\vk}
\int_{-\infty}^{\infty} dk_0 \, R
\,\left[\overline{\Sigma}_W(\vk,k_0)+\overline{\Sigma}_Z(\vk,k_0)\right]
 \;  L  \;  e^{i \vk\cdot(\vx-\vx')} \; e^{-i k_0 (t-t')} \;  \Theta(t-t'),
\ee
The contributions from charged and neutral vector bosons are given
by \be \overline{\Sigma}_W(\vk,k_0) = \frac{g^2}{2} \int \dbarq \int
dp_0 \int dq_0  \; \delta(p_0+q_0-k_0) \; \gamma^\mu  \;
\rho^f(\vk-\vq,p_0)  \; \rho^W_{\mu \nu}(\vk)(\vq,q_0) \; \gamma^\nu
\; [1-N_f(p_0)+N_b(q_0)] \; ,\label{retSWdisp} \ee \be
\overline{\Sigma}_Z(\vk,k_0) = \frac{g^2}{4 \; \cos^2\theta_w} \int
\dbarq \int dp_0 \int dq_0  \; \delta(p_0+q_0-k_0) \; \gamma^\mu \;
\rho^f(\vk-\vq,p_0) \; \rho^Z_{\mu \nu}(\vq,q_0) \; \gamma^\nu  \;
[1-N_f(p_0)+N_b(q_0)] \; ,\label{retSZdisp} \ee \noindent where $
\rho_{W,Z}(\vq,q_0) $ are the vector boson spectral densities given
by eq.(\ref{vbspecdens}) with $M\equiv M_{W,Z}$ respectively. It is
clear that $ \overline{\Sigma}_{W,Z}(\vk,k_0) $ corresponds to a
vector-like theory.

Using the integral representation of the function $ \Theta(t-t') $,
the retarded self-energy can be written in the following simple
dispersive form \be \Sigma_{ret}(\vx-\vx',t-t')=
\frac{1}{V}\sum_{\vk} \int_{-\infty}^{\infty} \frac{d\omega}{2\pi}
\; e^{i \vk\cdot(\vx-\vx')} \;  e^{-i\omega(t-t')} \;  R \; \left[
\Sigma_W(\vk,\omega)+ \Sigma_Z(\vk,\omega) \right] \;  L \; ,
\label{retSE2} \ee \be \Sigma_{W,Z}(\vk,\omega) = \int dk_0 \;
\frac{\overline{\Sigma}_{W,Z}(\vk,k_0)}{k_0-\omega-i\epsilon},
\label{retSEfin} \ee \noindent where $ \epsilon \rightarrow 0^+ $.
Hence, from the above expression, we identify \be
\overline{\Sigma}_{W,Z}(\vk,\omega) = \frac{1}{\pi} \;
\textit{Im}\,\Sigma_{W,Z}(\vk,\omega) \; .\label{imparts} \ee

Furthermore, since $ R \,(\pm m_f)\, L=0 $, the factor $ \pm m_f $
in the free fermionic spectral density defined in Eqns.
(\ref{rhofree}) and (\ref{ppm}) can be ignored when we compute $
R\,\left[\overline{\Sigma}_W(\vk,\omega)+
\overline{\Sigma}_Z(\vk,\omega) \right]\, L $. The signature of the
the fermion mass $ m_f $ is only reflected in the factors of $\p$ in
the spectral density. Ignoring the factor $ m_f $ from now on, the
fermionic spectral density is proportional to the $\gamma$ matrices
only and does not feature the identity matrix or $ \gamma^5 $.
Therefore, there is the following simplification \be
R\,\left[\overline{\Sigma}_W(\vk,\omega)+
\overline{\Sigma}_Z(\vk,\omega) \right]\, L = \left[
\overline{\Sigma}_W(\vk,\omega)+ \overline{\Sigma}_Z(\vk,\omega)
\right]\, L \; . \label{iden} \ee

\end{document}